\documentclass[aps,preprint,showpacs,groupedaddress,floatfix]{revtex4}
\usepackage{graphicx}
\usepackage{bm}
\usepackage{amsmath,amssymb}
\usepackage{times,txfonts}
\usepackage{eucal}

\renewcommand{\vec}[1]{\ensuremath{\bm{#1}}}

\newcommand{\eq}[1]{\begin{equation}#1\end{equation}}
\newcommand{\eqmulti}[1]{\begin{equation}\begin{split}#1\end{split}\end{equation}}

\newcommand{\bra}[1]{\ensuremath{\langle{#1}|\,}}
\newcommand{\ket}[1]{\ensuremath{\,|{#1}\rangle}}

\newcommand{\matrixe}[3]{\ensuremath{\langle{#1}|\,{#2}\,|{#3}\rangle}}

\newcommand{\op}[1]{\ensuremath{\mathrm{#1}}}
\newcommand{\adj}[1]{\ensuremath{{{#1}}^{\dag}}}
\newcommand{\corr}[1]{\ensuremath{\widetilde{#1}}}

\newcommand{\ii}{\ensuremath{\mathrm{i}}}
\newcommand{\dd}{\ensuremath{\mathrm{d}}}

\renewcommand{\vec}[1]{\ensuremath{\bm{#1}}}

\newcommand{\cO}{\ensuremath{\op{c}}}
\newcommand{\ccO}{\ensuremath{\adj{\op{c}}}}
\newcommand{\gO}{\ensuremath{\op{g}}}

\newcommand{\qO}{\ensuremath{\op{q}}}
\newcommand{\rO}{\ensuremath{\op{r}}}

\newcommand{\vO}{\ensuremath{\op{v}}}
\newcommand{\xO}{\ensuremath{\op{x}}}
\newcommand{\aO}{\ensuremath{\op{a}}}

\newcommand{\CO}{\ensuremath{\op{C}}}
\newcommand{\CCO}{\ensuremath{\adj{\op{C}}}}

\newcommand{\QO}{\ensuremath{\op{Q}}}

\newcommand{\HO}{\ensuremath{\op{H}}}

\newcommand{\TO}{\ensuremath{\op{T}}}
\newcommand{\VO}{\ensuremath{\op{V}}}

\newcommand{\XO}{\ensuremath{\op{X}}}

\newcommand{\AO}{\ensuremath{\op{A}}}

\newcommand{\PsiO}{\ensuremath{\op{\Psi}}}

\newcommand{\rV}{\ensuremath{\vec{r}}}

\newcommand{\JC}{\ensuremath{\mathcal{J}}}
\newcommand{\MC}{\ensuremath{\mathcal{M}}}
\newcommand{\QC}{\ensuremath{\mathcal{Q}}}
\newcommand{\TC}{\ensuremath{\mathcal{T}}}

\newcommand{\pOV}{\ensuremath{\vec{\op{p}}}}
\newcommand{\qOV}{\ensuremath{\vec{\op{q}}}}
\newcommand{\rOV}{\ensuremath{\vec{\op{r}}}}

\newcommand{\xOV}{\ensuremath{\vec{\op{x}}}}

\newcommand{\LOV}{\ensuremath{\vec{\op{L}}}}

\newcommand{\XOV}{\ensuremath{\vec{\op{X}}}}

\newcommand{\sigmaOV}{\ensuremath{\vec{\op{\sigma}}}}

\newcommand{\tensorO}{\ensuremath{\op{s}_{12}}}

\newcommand{\Rm}{\ensuremath{R_-}}

\newcommand{\Rp}{\ensuremath{R_+}}

\newcommand{\Rpm}{\ensuremath{R_{\pm}}}

\newcommand{\UCOM}{\ensuremath{\textrm{UCOM}}}
\newcommand{\cm}{\ensuremath{\textrm{cm}}}

\newcommand{\pmS}{\ensuremath{\!\pm\!}}
\newcommand{\mpS}{\ensuremath{\!\mp\!}}
\newcommand{\intr}{\ensuremath{\textrm{int}}}

\begin{document}

%=========================================================================
%  Header
%
\title{Collective multipole excitations based on correlated realistic nucleon-nucleon interactions}

\author{N. Paar\footnote{on leave of absence from Physics Department, Faculty of Science,
University of Zagreb, Croatia}\footnote{Electronic address: nils.paar@physik.tu-darmstadt.de}}
\author{P. Papakonstantinou}
\author{H. Hergert}
\author{R. Roth}

\affiliation{ Institut f\" ur Kernphysik, Technische Universit\" at Darmstadt, Schlossgartenstrasse 9,
D-64289 Darmstadt, Germany}

\date{\today}

\begin{abstract}
We investigate collective multipole excitations for closed shell nuclei from ${}^{16}$O to ${}^{208}$Pb using correlated realistic nucleon-nucleon interactions in the framework of the random phase approximation (RPA). The dominant short-range central and tensor correlations are treated explicitly within the Unitary Correlation Operator Method (UCOM), which provides a phase-shift equivalent correlated interaction $\VO_{\UCOM}$ adapted to simple uncorrelated Hilbert spaces. The same unitary transformation that defines the correlated interaction is used to derive correlated transition operators. Using $\VO_{\UCOM}$ we solve the Hartree-Fock problem and employ the single-particle states as starting point for the RPA. By construction, the UCOM-RPA is fully self-consistent, i.e. the same correlated nucleon-nucleon interaction is used in calculations of the HF ground state and in the residual RPA interaction. Consequently, the spurious state associated with the center-of-mass motion is properly removed and the sum-rules are exhausted within $\pm$3$\%$.  The UCOM-RPA scheme results in a collective character of giant monopole, dipole, and quadrupole resonances in closed-shell nuclei across the nuclear chart. For the isoscalar giant monopole resonance, the resonance energies are in agreement with experiment hinting at a reasonable compressibility. However, in the $1^-$ and $2^+$ channels the resonance energies are overestimated due to missing long-range correlations and three-body contributions.
\end{abstract}


\maketitle

%=========================================================================
%  Section 1
%
\section{\label{secI}Introduction}

One of great challenges for nuclear theory is the description of ground state properties and excitation phenomena in finite nuclei, based on realistic nucleon-nucleon (NN) interactions. A variety of highly accurate realistic NN potentials are presently available, e.g. Nijmegen~\cite{Sto.93}, CD Bonn~\cite{Mac.01}, and Argonne V18~\cite{Wir.95}. Recently, realistic NN potentials have been constructed in the framework of chiral perturbation theory, assuming a chiral symmetry breaking scale chosen as to maintain pions and nucleons as relevant degrees of freedom \cite{Ent.03}. However, a quantitative description of nuclear structure properties necessitates the inclusion of additional ingredients besides the two-nucleon interaction. In particular, it has been noted that relativistic corrections and three-nucleon (3N) interactions in light nuclear systems can give sizable contributions~\cite{Car.98}. Most of the available 3N interactions are phenomenological, i.e., their parameters are adjusted to experimental data in finite nuclei~\cite{Car.83,Pie.01,Nog.04}. So far, only the chiral approaches offer a consistent derivation of two- and three-nucleon forces.

These advances strengthend the interest in \emph{ab initio} nuclear structure calculations based on realistic potentials. Recent studies include different theoretical approaches for the description of ground state properties and low-lying excitation spectra in finite nuclei, e.g. Green's function Monte Carlo~\cite{Pie.04}, no-core shell model~\cite{Nav.05}, and coupled cluster method~\cite{Wlo.05}. Because of huge computational requirements, most of these methods are limited to light nuclei. In order to access heavier nuclei one has to resort to approximate solutions of the nuclear many-body problem. A particularly appealing method is the self-consistent Hartree-Fock (HF) scheme, which has been very successful in the description of various nuclear ground state properties. However, those calculations were usually based on simple phenomenological interactions such as Gogny or Skyrme interactions~\cite{Vau.72,Dec.80}, whose microscopic foundation is not well established, since their parameters are adjusted to the bulk nuclear properties. The use of bare realistic NN interactions in HF dramatically fails, resulting in unbound nuclei~\cite{Mut.00}. This is a direct consequence of strong many-body correlations induced by the short-range repulsion and the tensor part of the realistic potentials. The dominant short-range correlations cannot be described by many-body states given by a simple Slater determinant as in the HF scheme. Therefore, for practical applications of the HF approximation, the realistic NN interaction has to be converted into an effective interaction, which is adapted to the many-body model space under consideration. 

This objective can be achieved within the Unitary Correlation Operator Method (UCOM), which describes short-range central and tensor correlations explicitly by means of a unitary transformation \cite{Fel.98,Nef.03,Rot.04,Rot.05,RPPH.05}. Other methods employing unitary transformations have been developed, including the unitary model operator
approach~\cite{Pro.64,Suz.87,Fuj.04} and the Lee-Suzuki transformation~\cite{SL.80}. In contrast to these methods, the correlation operators in the UCOM approach are given explicitly, allowing for the derivation of a state-independent operator form of the correlated interaction and of other relevant correlated operators. Although different by its construction, the effective NN interaction from the UCOM method is similar to the $V_{low-k}$ potential obtained by using renormalization group concepts \cite{Bog.03}. Both approaches provide a phase-shift equivalent low-momentum interaction, which is appropriate for nuclear structure calculation in simple model spaces \cite{RPPH.05,Cor.03,Cor.05}. First applications of the correlated interaction in Hartree-Fock and many-body perturbation theory from ${}^{4}$He to ${}^{208}$Pb show the potential of this approach~\cite{RPPH.05}.

In the present work we study collective excitation phenomena in atomic nuclei using a correlated interaction based on the Argonne V18 potential. A particularly convenient method to investigate low-amplitude excitation phenomena is the random-phase approximation (RPA). Different versions of RPA and quasiparticle RPA, based on phenomenological interactions, have been very successful not only for the description of giant resonances and low-lying states (e.g. Refs. \cite{Row.70,BT.75,Spe.81,Dum.83,Daw.90,Ham.97,Col.00,Rin.01,Rod.02,Agr.03}), but also in studies of exotic nuclear structure of collective excitations in nuclei away from the valley of $\beta$-stability~\cite{Mat.01,Ter.04,Paa.03,Col.03,Sar.04,Paa_pp.05,Cao.05,Gor.05,Per.05}. In the present study, correlated realistic NN interactions are employed for the first time to investigate collective excitations in both light and heavy closed-shell nuclear systems. This serves as a stringent test of the UCOM framework and provides direct information about the physical properties of the underlying correlated realistic NN interactions.

In Sec.~\ref{secII} we introduce the basic formalism of the random-phase approximation in the framework of the unitary correlation operator method. In Sec.~\ref{secIII} we present some fundamental tests of validity of the scheme: the separation of the spurious center-of-mass motion and the accuracy of the sum rules in comparison to the standard expressions. Section~\ref{secIV} presents the application of the UCOM-RPA approach for the description of collective excitation phenomena, in particular giant monopole, dipole, and quadrupole resonances. The role of the range of the tensor correlator and the impact of missing long-range correlations and three-body forces is discussed. Finally, in Sec. \ref{secV} we summarize our findings.

%=========================================================================
%  Section 2

\section{\label{secII}The random-phase approximation based on the Unitary Correlation Operator
Method}

We employ correlated NN interactions constructed within the UCOM approach, for the description of small-amplitude oscillations around the nuclear ground state. In a first step we solve the HF equations based on the two-body matrix elements of the correlated realistic NN interaction. In a second step, the RPA equations are formulated in the HF single-nucleon basis.
Here we outline the basic principles of the UCOM scheme \cite{Fel.98,Nef.03,Rot.04,Rot.05} and the HF and RPA formalism.

\subsection{\label{ssecII-1}Unitary correlation operator method (UCOM)}

The Unitary Correlation Operator Method provides an effective NN interaction which can be directly used in nuclear structure calculations~\cite{Rot.04,Rot.05,RPPH.05}. The central idea of this approach is the explicit treatment of the interaction-induced short-range central and tensor correlations. These correlations are imprinted into an uncorrelated many-body state $\ket{\Psi}$ through a state independent unitary transformation defined by the unitary correlation operator $\CO$, resulting in a correlated state $\ket{\corr{\PsiO}}$,
\eq{ \label{eq:corr_state}
  \ket{\corr{\PsiO}} = \CO\; \ket{\Psi} \;.
}
Even for the simplest uncorrelated state, a Slater determinant, the correlated state $\ket{\corr{\PsiO}}$ contains the relevant short-range correlations. Any expansion of $\ket{\corr{\PsiO}}$ in a basis of Slater determinants will require a huge number of basis states, i.e. a large model space. Hence, the unitary transformation reduces the size of the model space necessary for an adequate respresentation of the full many-body state.

The correlation operator $\CO$ is written as a product of unitary operators $\CO_{\Omega}$ and $\CO_{r}$ describing tensor and central correlations, respectively. Both are formulated as exponentials of a Hermitian generator, 
\eq{ \label{eq:correlator}
  \CO = \CO_{\Omega} \CO_{r}
  = \exp\!\Big[-\ii \sum_{i<j} \gO_{\Omega,ij} \Big]\;
    \exp\!\Big[-\ii \sum_{i<j} \gO_{r,ij} \Big]\;.
}
The two-body generators $\gO_r$ and $\gO_{\Omega}$ are constructed following the physical mechanisms by which the interaction induces central and tensor correlations. The short-range central correlations, caused by the repulsive core of the interaction, are introduced by a radial distance-dependent shift pushing nucleons apart from each other if they are within the range of the core. Radial shifts are generated by the component of the relative momentum $\qOV  = \frac{1}{2} [\pOV_1 - \pOV_2]$ along the distance vector $\rOV = \xOV_1 - \xOV_2$ of two particles \cite{Rot.04},
\eq{
  \qO_r 
  = \frac{1}{2} [ \qOV\cdot\tfrac{\rOV}{\rO} 
    + \tfrac{\rOV}{\rO}\cdot\qOV ] \;.
}
The dependence of the radial shift on the particle distance is described by a function $s(r)$ in the Hermitian generator,

\eq{ \label{eq:central_generator}
  \gO_r 
  = \frac{1}{2} [ s(\rO) \qO_r + \qO_r s(\rO) ] \;.
}

The application of $\cO_{r}$ in two-body space corresponds to a norm conserving coordinate transformation $\rV \mapsto \Rm(r) \frac{\rV}{r}$ with respect to the relative coordinate. The radial correlation function $\Rm(r)$ and its inverse $\Rp(r)$ are related to the function $s(r)$ in the following way \cite{Rot.04},
\eq{
  \int_r^{\Rpm(r)} \frac{\dd\xi}{s(\xi)} = \pm 1. 
}
For a given bare potential, 
the central correlation functions $R_{+}(r)$ are determined by an energy minimization in the two-body system for each $(S,T)$ channel. In the purely repulsive channel $(S,T)=(0,0)$, an explicit constraint on the range of the central correlator is used,
\eq{ \label{central_constraint}
  \int\dd{r}\; r^2 (\Rp(r)-r) = I_{R_{+}}. 
}
Throughout this work, we use the Argonne V18 potential with the optimal correlators constructed in Ref.~\cite{Rot.05}.
We adopt the short-range central correlator with the constraint $I_{R_{+}}^{(S=0,T=0)}=0.1$ fm$^4$. 
We have verified that variations around this value have negligible effect on the ground state properties and excitation spectra in $0^+$, $1^-$, and $2^+$ channels, in closed-shell nuclei across the nuclear chart.

Tensor correlations between two nucleons are generated by a spatial shift perpendicular to the radial direction. In practice, this can be achieved by using the ``orbital momentum'' operator
\eq{
  \qOV_{\Omega} 
  = \qOV - \tfrac{\rOV}{\rO}\;\qO_r
  = \frac{1}{2\rO^2} [\,\LOV\times\rOV - \rOV\times\LOV \,] \;,
}
where $\LOV$ is the relative orbital angular momentum operator. Radial momentum $\tfrac{\rOV}{\rO}\qO_r$ and orbital momentum $\qOV_{\Omega}$ constitute a special decomposition of the relative momentum operator $\qOV$ and generate shifts orthogonal to each other. The dependence of the shift on the spin orientation is implemented in the following way in the generator $\gO_{\Omega}$ \cite{Nef.03}: 
\eq{ \label{eq:tensorgen}
  \gO_{\Omega}  
  = \frac{3}{2}\vartheta(\rO)
    \big[(\sigmaOV_1\!\cdot\qOV_{\Omega})(\sigmaOV_2\!\cdot\rOV) 
    + (\sigmaOV_1\!\cdot\rOV)(\sigmaOV_2\!\cdot\qOV_{\Omega}) \big].
}
The two spin operators and the relative coordinate $\rOV$ enter in a similar manner like in the standard tensor operator $\tensorO$, but one of the coordinate operators is replaced by the orbital momentum $\qOV_{\Omega}$, which generates the transverse shift. The size and radial dependence are given by a tensor correlation function $\vartheta(r)$ for each of the two $S=1$ channels. The parameters of $\vartheta(r)$ are determined from an energy minimization in the two-body system \cite{Rot.05}, with an additional restriction on the correlation volume which constrains the range of the tensor correlator, 
\eq{ \label{tensor_constraint}
  \int\dd{r}\; r^2 \vartheta(r) = I_{\vartheta}.
}
As for the central correlators we adopt the optimal correlation functions for the Argonne V18 potential determined in Ref.~\cite{Rot.05}. Applications within the no-core shell model have show that $I_{\vartheta}^{(S=1,T=0)}$ = 0.09 fm$^3$ leads to the best description of binding energies in $^{3}$H and $^{4}$He~\cite{Rot.05}. This correlator set \footnote{No tensor correlator is employed in the $(S=1, T=1)$ channel, since the tensor interaction is much weaker there.}, which we refer to as standard correlator in the following, also provides a good description of binding energies for heavier nuclei within many-body perturbation theory \cite{RPPH.05}. In addition to the standard correlator, we will employ other values for the constraint on the range of the tensor correlator, $I_{\vartheta}^{(S=1,T=0)}$ = 0.07, 0.08, and 0.09 fm$^3$, in order to probe its relevance for the description of the global properties of collective excitation phenomena in atomic nuclei.

%%%%%%%%%%%%%%%%%%%%%%%%%%%%%%%%%%%%%%%%%%%%%%%%%%%%%%%%%%%%%%%%%%%%%%%%%%%%%%%
\subsection{\label{ssecII-ope}Correlated operators}

Due to the unitarity of the correlation operator, matrix elements of an operator $\AO$ with correlated many-body states are equal to those evaluated with the correlated operator $\corr{\AO}$ and uncorrelated many-body states, i.e. 
\begin{equation}
\matrixe{\corr{\PsiO}}{\AO}{\corr{\PsiO'}}=
\matrixe{\Psi}{\CCO \AO \,\CO}{\Psi'}= 
\matrixe{\Psi}{\corr{\AO}}{\Psi'}.
\end{equation}
The correlated operator contains irreducible contributions to all particle numbers. Within a cluster expansion of the correlated operator
\eq{
  \corr{\AO}
  = \CCO \AO \CO
  = \corr{\AO}^{[1]} + \corr{\AO}^{[2]} + \corr{\AO}^{[3]} + \cdots \;,
\label{cexpansion}
}
where $\corr{\AO}^{[n]}$ denotes the irreducible $n$-body contribution, we usually assume a two-body approximation, i.e. three-body and higher-order terms of the expansion are neglected. In previous studies it has been verified that higher order contributions due to short-range central correlations can be neglected in the description of nuclear structure properties \cite{Rot.04}. However, this is not the case for the tensor correlations. The tensor interaction is very long-ranged and thus generates long-range tensor correlations in an isolated two-body system, e.g., the deuteron. In a many-body system, the long-range tensor correlations between a pair of nucleons are suppressed by the presence of other nucleons, leading to a screening of the tensor correlations at large interparticle distances. In terms of the cluster expansion this screening appears through significant higher-order contributions. In order to avoid large higher-order contributions we have restricted the range of the tensor correlation function (cf. Sec. \ref{ssecII-1}), which provides an effective inclusion of the screening effect \cite{Rot.05}. 

Starting from the uncorrelated Hamiltonian for the $A$-body system,
\eq{ 
  \HO 
  = \TO + \VO
  = \sum_{i=1}^{A} \frac{1}{2 m_N}\, \pOV_i^2
    + \sum_{i>j=1}^{A} \vO_{ij} \;,    
}
consisting of the kinetic energy operator $\TO$ and a two-body potential, the formalism of the unitary correlation operator method is employed to construct the correlated Hamiltonian in two-body approximation
\begin{equation}
  \HO^{C2} = {\corr{\TO}}^{[1]} + {\corr{\TO}}^{[2]} + {\corr{\VO}}^{[2]}
  = \TO + \VO_{\UCOM},
\end{equation}
where the one-body contribution comes only from the uncorrelated kinetic energy ${\corr{\TO}}^{[1]} = \TO$.
Two-body contributions arise from the correlated kinetic energy ${\corr{\TO}}^{[2]}$ and the correlated potential ${\corr{\VO}}^{[2]}$, which together constitute the phase-shift equivalent correlated interaction $\VO_{\UCOM}$ ~\cite{Rot.05}.

The correlated realistic NN interaction $\VO_{\UCOM}$ is a good starting point for a study of nuclear structure. However, one should keep in mind that long-range correlations and residual three-body forces are not yet included. One can account for long-range correlations by a suitable extension for the many-body space as discussed in Ref.~\cite{RPPH.05}. The problem of effective three-body interactions, which are composed of the genuine three-body force and the three-body terms of the cluster expansion, remains as an objective for future studies. Therefore, in the present work we focus on the question to which extent the correlated NN interaction alone is sufficient for the description of collective excitations in finite nuclei.

%%%%%%%%%%%%%%%%%%%%%%%%%%%%%%%%%%%%%%%%%%%%%%%%%%%%%%%%%%%%%%%%%%%%%%%%%%%%
\subsection{\label{ssecII-2a}Hartree-Fock method with correlated realistic
NN-interactions}

The correlated realistic NN potential $\VO_{\UCOM}$ is suitable for the use in HF calculations for the ground state of finite nuclei~\cite{RPPH.05}. We start from a Hamiltonian which consists of kinetic energy and the $\VO_{\UCOM}$ interaction derived from the Argonne V18 potential including the Coulomb potential \footnote{The charge-dependent terms of the correlated Argonne V18 potential have a negligible effect and are omitted furtheron.}. More details about the UCOM-HF scheme are available in Ref.~\cite{RPPH.05}. The center-of-mass contributions are subtracted on the operator level, i.e. we employ the correlated intrinsic Hamiltonian in two body approximation, 
\eqmulti{\label{eq:hintr}
  \corr{\HO}_{\intr} 
  = \TO - \TO_{\cm} + \VO_{\UCOM} 
  = \TO_{\intr} + \VO_{\UCOM} \;.
}
The intrinsic kinetic energy operator reads,
\eq{\label{intrke}
  \TO_{\intr} 
  = \TO - \TO_{\cm} 
  = \frac{2}{A} \frac{1}{m_N} \sum_{i<j}^{A} \qOV_{ij}^2 \;,
}
where $\qOV$ corresponds to the relative two-body momentum operator, and we assume equal proton and neutron masses and thus a reduced mass $\mu=m_N/2$. Assuming spherical symmetry, the HF single-particle states are expanded in a basis of harmonic oscillator eigenstates,
\eq{  
  \ket{\nu l j m m_t} 
  = \sum_n C^{(\nu l j m m_t)}_{n} \ket{n l j m m_t} \;,
}
with radial quantum number $n$, orbital angular momentum $l$, total angular momentum $j$ with projection $m$, and isospin projection quantum number $m_t$. For closed shell calculations, we restrict $C^{(\nu l j m m_t)}_{n}$ to be independent of $m$. The HF equation can be written in matrix form,
\eq{ \label{eq:hf_hfeigenproblem}
  \sum_{\bar{n}} h^{(l j m_t)}_{n\bar{n}} C^{(\nu l j m_t)}_{\bar{n}} 
  = \epsilon^{(\nu l j m_t)} C^{(\nu l j m_t)}_{n} \;,
}
which is solved self-consistently to determine the expansion coefficients and single-particle energies. The single-nucleon Hamiltonian,
\eq{ \label{eq:hf_hfhamiltonian}
  h^{(l j m_t)}_{n\bar{n}}
  = \sum_{l',j',m'_t} \sum_{n',\bar{n}'} 
    H^{(l j m_t l' j' m'_t)}_{n n'; \bar{n} \bar{n}'}
    \varrho^{(l' j' m'_t)}_{n' \bar{n}'}, 
}
is constructed from the $m$-averaged antisymmetric two-body matrix elements of the correlated intrinsic Hamiltonian $\corr{\HO}_{\intr}$ \eqref{eq:hintr},
\eqmulti{ \label{eq:hf_twobodyme}
  &H^{(l j m_t l' j' m'_t)}_{n n', \bar{n} \bar{n}'} 
  = \frac{1}{(2j+1)(2j'+1)} \sum_{m,m'} \\
  &\quad\times \matrixe{n l j m m_t, n' l' j' m' m'_t}
    {\corr{\HO}_{\intr}}{\bar{n} l j m m_t, \bar{n}' l' j' m' m'_t},
}
where $\varrho^{(l' j' m'_t)}_{n' \bar{n}'}$ corresponds to the one-body density matrix.

The harmonic oscillator basis is typically restricted to $13$ major shells, which warrants complete convergence of the HF results. Calculations with larger basis sizes are possible but rather time-consuming, because of the computational effort for evaluating the two-body matrix elements of the $\VO_{\UCOM}$ potential. The optimal value of the harmonic oscillator length $a_0=\sqrt{\hbar/m_{N}\omega_0}$ is determined from an explicit energy minimization for different regions in the nuclide chart.

\begin{figure}
\includegraphics[width=0.9\columnwidth]{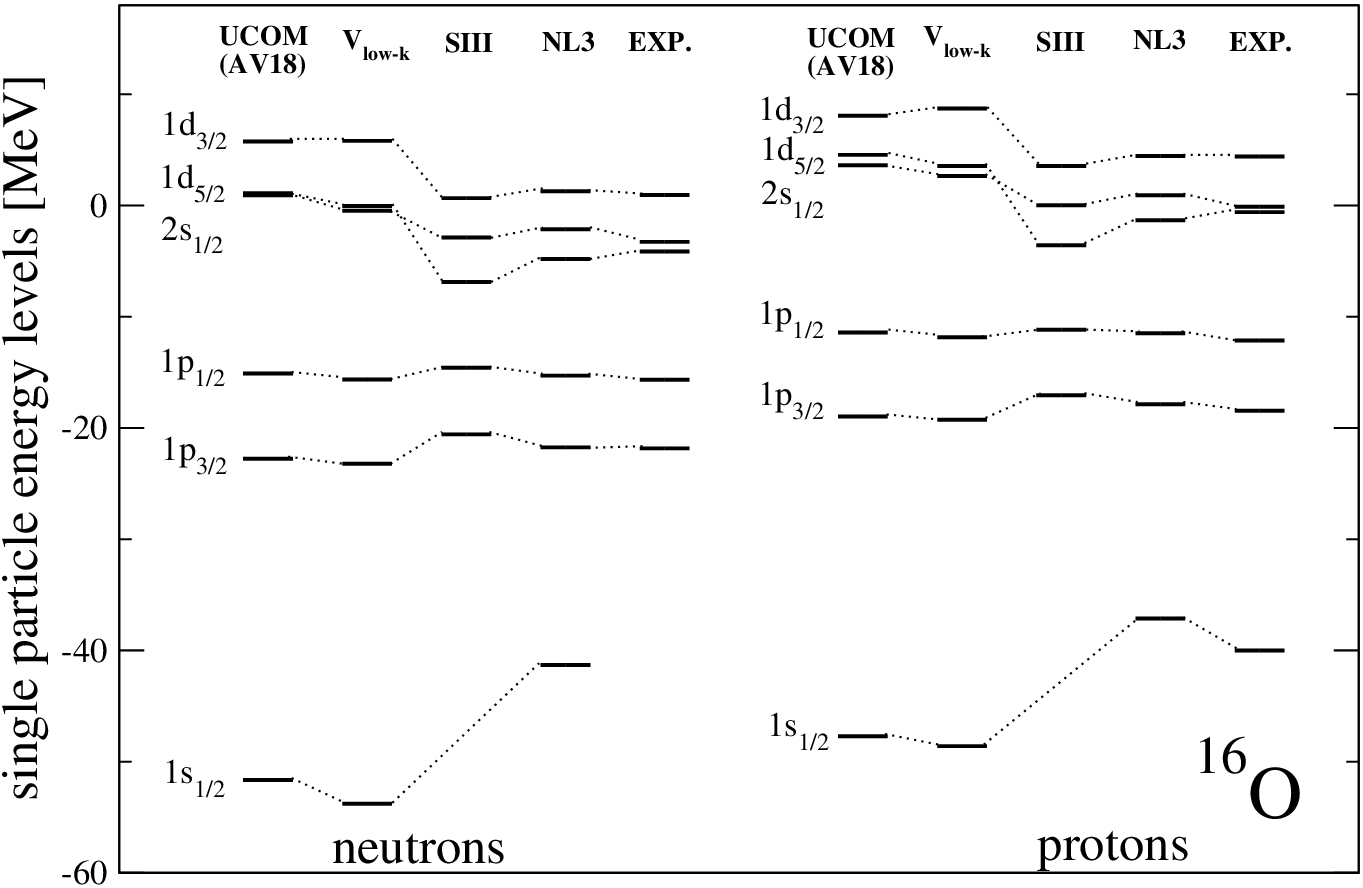}
\vspace{1cm}
\caption{The UCOM-HF neutron and proton single particle spectrum for $^{16}$O,
along with the corresponding HF spectra based on the
low-momentum NN potential $V_{low-k}$~\protect\cite{Cor.03}, phenomenological 
HF with the SIII Skyrme-type interaction~\protect\cite{Isa.02}, relativistic mean field theory
with the NL3 effective interaction~\protect\cite{LKR.97}, and the experimental spectra~\cite{Isa.02}. 
The UCOM-HF calculations are based on the correlated Argonne
V18 interaction with the standard
constraints on the ranges of the central and tensor correlators~(Sec.~\ref{ssecII-1}).
}
\label{figspectra1}
\end{figure}

\begin{figure}
\includegraphics[width=0.9\columnwidth]{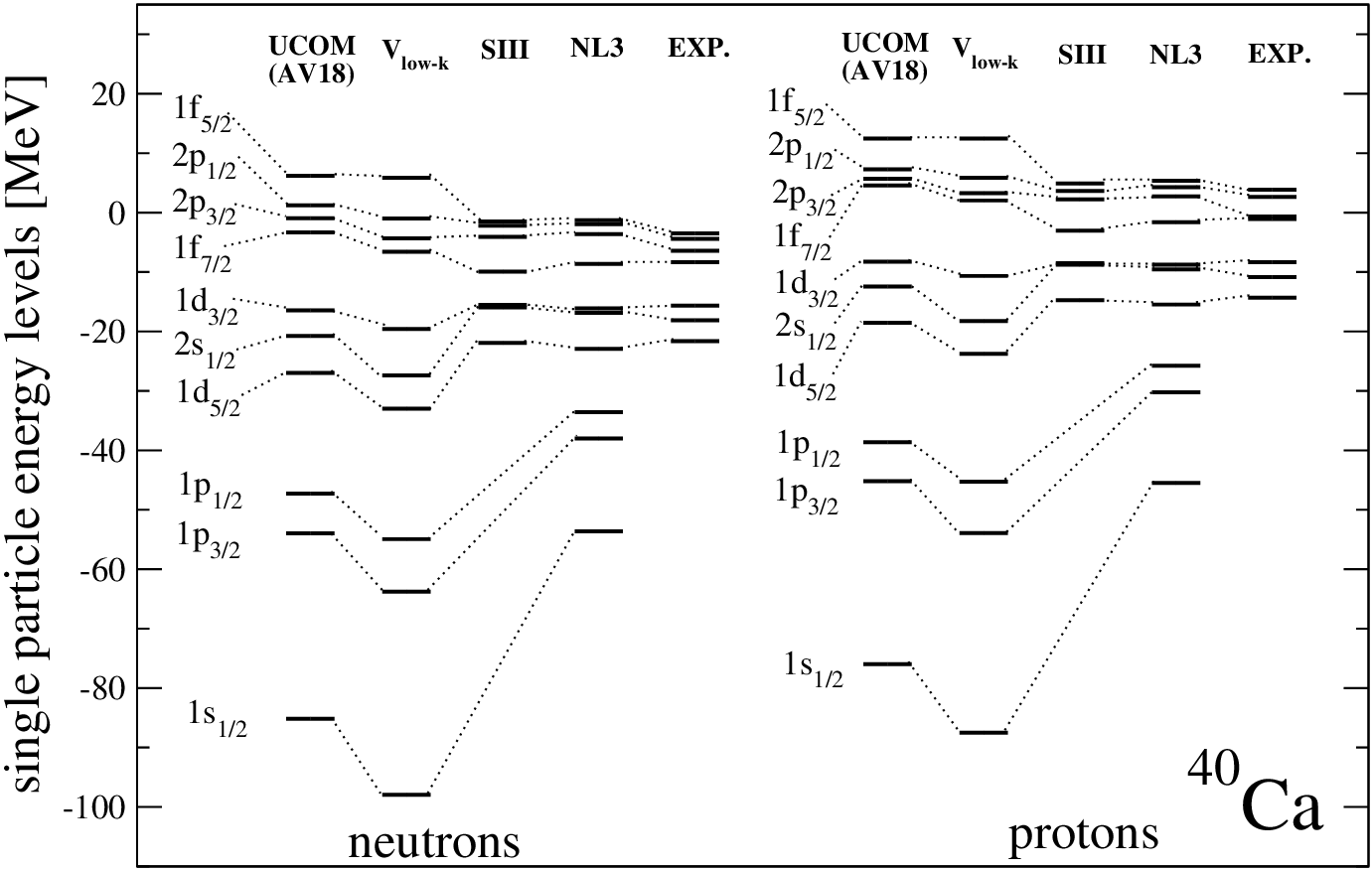}
\vspace{1.5cm}
\caption{The same as Fig.~\ref{figspectra1}, but for $^{40}$Ca.}
\label{figspectra2}
\end{figure}

In Figs. \ref{figspectra1} and \ref{figspectra2} we show the UCOM-HF single-nucleon spectra near the Fermi level for $^{16}$O and $^{40}$Ca, respectively. The calculations are based on the correlated Argonne V18 interaction, using the standard
correlator (Sec.\ref{ssecII-1}). The calculated energy levels are compared with the HF spectra obtained with the low-momentum NN potential $V_{low-k}$, with two standard phenomenological interactions in the nonrelativistic (Skyrme) \cite{Isa.02} and the relativistic (NL3) \cite{LKR.97} framework, and with experimental levels which are evaluated from the binding energies \cite{Isa.02}. For the case of $^{16}$O, the UCOM-HF single-particle spectrum is similar to the one obtained with the $V_{low-k}$ potential. These two approaches, however, result in somewhat different spectra than the phenomenological models and those extracted from the experiment. For HF approaches based on realistic NN interactions, the spectra appear spread too wide in energy. The results for $^{40}$Ca are similar, but the UCOM-HF spectra for neutrons and protons are slightly more compressed in comparison to the $V_{low-k}$ case. 

The origin for this behavior is in the missing long-range correlations, higher order terms of the cluster expansion, and genuine three-body interactions. One way to account the long-range correlations for the binding energies and radii beyond the simple mean-field approach is the framework of the many-body perturbation theory (MBPT). When employed in the UCOM scheme, the second order MBPT recovers quite a significant part of the missing binding in UCOM-HF~\cite{RPPH.05}. An alternative approach to account for the long-range correlations in the ground state would be the inclusion of the correlations due to low-lying excited states and collective excitations within an RPA framework~\cite{Rei.85,Esb.83}.

%%%%%%%%%%%%%%%%%%%%%%%%%%%%%%%%%%%%%%%%%%%%%%%%%%%%%%%%%%%%%%%%%%%%%%%%%%%%%%%%%
\subsection{\label{ssecII-2}Outline of the UCOM random-phase approximation}

In the limit of small-amplitude oscillations of the nuclear density around the ground state, collective excitation phenomena can be studied within the random-phase approximation (RPA). We address the question to which extent are the UCOM-HF single-nucleon basis and the residual interaction based on the correlated realistic NN interaction sufficient for a description of highly collective excitation phenomena, such as giant resonances. Since details about the derivation of RPA equations are available in textbooks \cite{Row.70, Rin.80}, we review the basic principles only briefly. The UCOM-HF single-particle states are used for the construction of the particle-hole~($ph$) configuration space for the RPA scheme. Assuming a spherical nuclear system, the coupling to good angular momentum is employed. The collective excited states of multipolarity $\JC$ are generated by the quasiboson operator,
\begin{equation}
\QO^+_{\nu,\JC\MC}\ket{0} = \ket{\nu},
\end{equation}
where the RPA vacuum $\ket{0}$ is defined by the condition,
\begin{equation}
\QO_{\nu,\JC\MC}\ket{0} = 0. 
\end{equation}
The quasiboson operator reads,
\begin{equation}
\QO^+_{\nu,\JC\MC}=\sum_{ph} \left[ X_{ph}^{\nu,\JC\MC}{\AO_{ph}^{\JC\MC}}^{+} -Y_{ph}^{\nu,\JC\MC}(-1)^{\JC-\MC}\AO_{ph}^{\JC,-\MC} \right], 
\end{equation}
where the sum runs over the $ph$ states of the HF single-nucleon basis, and
\begin{equation}
{\AO_{ph}^{\JC\MC}}^{+}=\sum_{m_p m_h}\bra{j_p m_p,j_h m_h}\JC\MC \rangle (-1)^{j_h-m_h}
\aO^{+}_{j_p m_p}\aO_{j_h,m_h}
\end{equation}
corresponds to the $ph$ creation operator. One of the standard
approaches to derive RPA
is the equation of motion method using the quasiboson 
approximation \cite{Row.70}, with the RPA vacuum approximated 
by the HF ground state, i.e. $\ket{0} \approx \ket{HF}$. The resulting 
set of coupled equations for the amplitudes $X_{ph}^{\nu,\JC\MC}$ and $Y_{ph}^{\nu,\JC\MC}$ 
and the RPA eigenvalues $\omega_{\nu}$ is given by,
%
% RPA EQUATIONS
\begin{equation}
\label{rpaeq}
\left(
\begin{array}{cc}
A^{\JC} & B^{\JC} \\
B^{\JC \ast} & A^{\JC \ast}
\end{array}
\right)
\left(
\begin{array}{c}
X^{\nu,\JC\MC} \\
Y^{\nu,\JC\MC}
\end{array}
\right) =\omega_{\nu}\left( 
\begin{array}{cc}
1 & 0 \\
0 & -1
\end{array}
\right)
\left( 
\begin{array}{c}
X^{\nu,\JC\MC} \\
Y^{\nu,\JC\MC}
\end{array}
\right)\; .
\end{equation}
The RPA matrices for the given configuration space of the single-nucleon
UCOM-HF basis are obtained from,
\begin{equation}
\begin{split}
A^{\JC}_{php'h'} 
&= \bra{HF}\big[ \big[\AO_{ph}^{\JC\MC},\corr{\HO}_{\intr}\big],{\AO_{p'h'}^{\JC\MC}}^{+} \big] \ket{HF} \\
B^{\JC}_{php'h'} 
&= -\bra{HF}\big[ \big[\AO_{ph}^{\JC\MC},\corr{\HO}_{\intr}\big],(-1)^{\JC-\MC}{\AO_{p'h'}^{\JC,-\MC}} \big] \ket{HF},
\end{split}
\end{equation}
where the Hamiltonian $\corr{\HO}_{\intr}$ includes the
intrinsic kinetic energy (\ref{intrke}) together 
with the correlated potential $\VO_{\UCOM}$, in a consistent way with the
Hartree-Fock equations.
In the present study, we assume that the RPA vacuum is rather well
approximated by the HF ground state. In principle, however, 
one would need to build excitations on the RPA vacuum, and iteratively
solve the extended RPA equations~\cite{Cat.98}. 
A preliminary study within an extended RPA framework indicates that the
proper implementation of the RPA vacuum causes only small
corrections in the excitation spectra. In particular, the centroid
energies of the strength distributions systematically decrease by
approx. 1 MeV in the isovector channel, while the excitation energies of
isoscalar modes are affected even less and increase by less than 1 MeV. 
More details will be provided in a forthcoming publication.

%%%%%%%%%%%%%%%%%%%%%%%%%%%%%%%%%%%%%%%%%%%%%%%%%%%%%%%%%%%%%%%%%%%%%%%%%%%%%%
\subsection{Transition operators}
The response for electric multipole transitions is given by the reduced transition probability~\cite{Rin.80},
\begin{equation}
B^{\TC}(E\JC, J_i \to J_f) = \frac{1}{2J_i+1}\big\vert \langle
f || \QC^{\TC}_{\JC} || i \rangle \big\vert^2,
\label{trgen}
\end{equation} 
where $\QC^{\TC}_{\JC}$ corresponds to the electric multipole transition operators. The isoscalar monopole operator is defined as
\begin{equation}
\QC^{\TC=0}_{00} = \sum_{i=1}^A \xO_i^2\; Y_{00}(\hat{\xOV}_i),
\label{monoperator}
\end{equation}
where $\xO_i=|{\xOV}_i|$. The multipole $(\JC>0)$ isoscalar $(\TC=0)$ and isovector $(\TC=1)$ operators are given by
\begin{equation}
\QC^{\TC=0}_{\JC\MC} = e\sum_{i=1}^{A}\xO_i^{\JC}\; Y_{\JC\MC}(\hat{\xOV}_i) 
\label{istroperator}
\end{equation}
and
\begin{equation}
\QC^{\TC=1}_{\JC\MC} = e\sum_{i=1}^{A}
\tau_z^{(i)}\xO_i^{\JC}\; Y_{\JC\MC}(\hat{\xOV}_i),
\label{ivtroperator}
\end{equation}
respectively.
In the UCOM framework, the operators of all observables need to be transformed in a consistent way.
Therefore, the same unitary transformation that is used for the nuclear Hamiltonian has to be employed for the multipole operators entering into transition matrix elements.
The effect of using correlated transition operators will be examined in the cases of monopole and quadrupole transitions.
Since the unitary correlation operators act on the relative coordinates, we rewrite the monopole and quadrupole transition operator in a two-body form
with separated relative and center-of-mass (c.m.) contributions (cf.~Ref.~\cite{Ste.05}),
\begin{equation}
\begin{split}
\QC^{\TC=0}_{00}  
&= \sum_{i}^A \xO_i^2 Y_{00}(\hat{\xOV}_i) \\
&= \frac{1}{2(A-1)} \sum_{i>j}^A
  \left[\rO_{ij}^2 Y_{00}(\hat{\rOV}_{ij}) + 4\XO_{ij}^2 Y_{00}(\hat{\XOV}_{ij})\right] 
\end{split}
\end{equation}
\begin{equation}
\begin{split}
\QC^{\TC=0}_{2\MC} 
&= e \sum_{i}^A \xO_i^2 Y_{2\MC}(\hat{\xOV}_i) \\ 
&= \frac{e}{2(A-1)} \sum_{i>j}^A  
  \left[\rO_{ij}^2 Y_{2\MC}(\hat{\rOV}_{ij}) + 4\XO_{ij}^2 Y_{2\MC}(\hat{\XOV}_{ij})\right],
\end{split}
\end{equation}
where $\rOV_{ij} = \xOV_i - \xOV_j$ are the relative and $\XOV_{ij} = (\xOV_i + \xOV_j)/2$ the c.m. coordinates of a nucleon pair. 

The correlated multipole operator is constructed in an analogous way as the correlated operators of the realistic NN interaction~\cite{Rot.05}. In two-body approximation we obtain
\begin{equation}
\label{corr_op}
\corr{\QC}^{C2} = \corr{\QC}^{[1]}+\corr{\QC}^{[2]} = \QC + \corr{\QC}^{[2]},
\end{equation}
where $\corr{\QC}^{[1]}=\QC$ corresponds to the bare one-body transition operator, and the two-body part of the correlated operator reads 
\begin{equation}
\label{corr_mono}
[\corr{\QC}^{\TC=0}_{00}]^{[2]}
= \frac{1}{2(A-1)}\sum_{i>j}^A 
  \big\{ \ccO\; [\rO_{ij}^2 Y_{00}(\hat{\rOV}_{ij})]\;\cO 
    - \rO_{ij}^2 Y_{00}(\hat{\rOV}_{ij}) \big\} 
\end{equation}
for the isoscalar monopole transition operator, and
\begin{equation}
\label{corr_quad}
[\corr{\QC}^{\TC=0}_{2\MC}]^{[2]}
= \frac{1}{2(A-1)}\sum_{i>j}^A 
  \big\{ \ccO\; [\rO_{ij}^2 Y_{2\MC}(\hat{\rOV}_{ij})]\;\cO 
    - \rO_{ij}^2 Y_{2\MC}(\hat{\rOV}_{ij}) \big\} 
\end{equation}
for the isoscalar quadrupole transition operator. Therefore, in addition to the usual transition matrix elements for the bare multipole operator,
\begin{equation}  
\matrixe{\nu}{{\QC}^{\TC}_{\JC \MC}}{0} =
\frac{1}{\hat{\JC}}
\sum_{ph} \big\{ X^{\nu,\JC}_{ph} \langle
p || \QC || h \rangle+(-1)^{j_{p}-j_{h}+\JC} \, Y^{\nu,\JC}_{ph}
\, \langle h || \QC || p \rangle \big\}, 
\label{tran1body}  
\end{equation} 
we also have to include contributions from the correlated two-body part,
\begin{equation}
\begin{split}  
\matrixe{\nu}{[\corr{\QC}^{\TC}_{\JC \MC}]^{[2]}}{0} 
&=\sum_{ph J_1 J_2 h'}(-1)^{j_p+j_{h'}}\sqrt{1+\delta_{hh'}}
  \left\{
  \begin{array}{ccc}
  j_p & j_h & \JC \\
  J_2 & J_1 & j_{h'}
  \end{array}
  \right\}
  \frac{\hat{J_1}\hat{J_2}}{\hat{\JC}} \\
&\times\Big\{ (-1)^{\JC+J_2}X^{\nu,\JC}_{ph} \langle (j_pj_{h'})J_1 || [\corr{\QC}^{\TC}_{\JC}]^{[2]} || (j_hj_{h'})J_2 \rangle \\
&\quad+(-1)^{J_1} Y^{\nu,\JC}_{ph} \langle (j_hj_{h'})J_2 || [\corr{\QC}^{\TC}_{\JC}]^{[2]} || (j_pj_{h'})J_1 \rangle
\Big\},
\label{tran2body}
\end{split}  
\end{equation} 
for each RPA eigensolution $( \omega_{\nu},X^{\nu},Y^{\nu})$, using the standard notation $\hat{\JC}=\sqrt{2\JC+1}$. The resulting transition strength function, including both the one and two-body contribution of the correlated transition operator, is given by
\begin{equation}
\begin{split}  
&B^{\TC}(E\JC, J_i \to J_f) 
= B^{\TC}_{\JC}(\omega_{\nu}) \\
&\qquad= \frac{1}{2J_i+1} \big| \langle \nu || \QC^{\TC}_{\JC} || 0 \rangle+
  \langle \nu || [\corr{\QC}^{\TC}_{\JC}]^{[2]}|| 0 \rangle \big| ^2.  
\label{strength}
\end{split}  
\end{equation} 

In order to evaluate the two-body matrix elements of the transition operator we use the relative spin-orbit coupled states of the form $\ket{n (L S) J M \; T M_T}$, with a generic radial quantum number $n$, relative orbital angular momentum $L$, spin $S$, total angular momentum $J$, and isospin $T$. In the following, the projection quantum numbers $M$ and $M_T$ are suppressed. For the two-body parts of the correlated monopole \eqref{corr_mono} and quadrupole operator \eqref{corr_quad} we have to evaluate general matrix elements of the form $\matrixe{n(L S)J T}{\ccO_{r} \ccO_{\Omega}\, f(\rO) Y_{\JC\MC}(\hat{\rO})\, \cO_{\Omega}\cO_{r}}{n'(L' S)J' T}$, where $\cO_{\Omega}$ and $\cO_{r}$ indicate the central and tensor correlation operators in two-body space. 

As for the correlated Hamiltonian  it is convenient to apply the tensor correlator to the two-body states and the central correlator to the operator \cite{Rot.05}. To this end we formally interchange the central and the tensor correlation operators using the identity $\cO_{\Omega}\cO_{r} = \cO_{r} \ccO_{r} \cO_{\Omega}\cO_{r} =  \cO_{r}\corr{\cO}_{\Omega}$. For the transformed tensor correlator we get
\eq{ \label{eq:tensorgentrans}
  \corr{\cO}_{\Omega}
  = \ccO_{r} \cO_{\Omega} \cO_{r}
  = \exp[-\ii \corr{\gO}_{\Omega}]
}
with 
\eq{
  \corr{\gO}_{\Omega}
  = \frac{3}{2}\,  \vartheta(\Rp(\rO))\, 
    \big[(\sigmaOV_1\!\cdot\qOV_{\Omega})(\sigmaOV_2\!\cdot\rOV) 
    + (\sigmaOV_1\!\cdot\rOV)(\sigmaOV_2\!\cdot\qOV_{\Omega}) \big]  .
}
The tensor correlation operator acts like the identity operator on the $LS$-coupled two-body states with $L=J$~\cite{Rot.05}, while for $L=J\pm1$ it gives 
\eqmulti{
  \corr{\cO}_{\Omega} \ket{n(J\mp 1,1) J T} 
  &= \cos\corr{\theta}_{J}(\rO)\, \ket{n (J\mp 1,1) J T } \\
  &\pm \sin\corr{\theta}_{J}(\rO)\, \ket{n (J\pm 1,1) J T} \;,
\label{tensorCtilde}
}
where
\eq{
  \corr{\theta}_{J}(\rO) 
  = 3 \sqrt{J(J+1)}\, \vartheta(\Rp(\rO)) \;.
 \label{thetatilde}
}   

Using Eq.~\eqref{tensorCtilde} we can easily derive explicit expressions for the correlated two-body matrix elements $\matrixe{n(L S)J T}{ \corr{\ccO}_{\Omega} \ccO_{r}\, f(\rO) Y_{\JC\MC}(\hat{\rOV})\, \cO_{r} \corr{\cO}_{\Omega} }{n'(L' S)J' T}$ of monopole and quadrupole transition operators. For $L=J$ and $L'=J'$ the tensor correlator is inactive and the correlated matrix element assumes the simple form 
\eqmulti{ \label{eq:corr_me_local1}
  &\matrixe{n(J S)J T}{\ccO_{r} \ccO_{\Omega}\, f(\rO) Y_{\JC\MC}(\hat{\rOV})\, 
    \cO_{\Omega}\cO_{r}}{n'(J' S)J' T} = \\ 
  &= \int\!\dd{\rO}\,u_{n,J}^{\star}(r)\, u_{n',J'}(\rO)\;
     \corr{f}(\rO)\; \matrixe{(J S) J T}{Y_{\JC\MC}(\hat{\rOV})}{(J'S)J' T},
}
where $\corr{f}(\rO) = f(\Rp(\rO))$ is the transformed radial dependence of the multipole operator, and $u_{n,L}(\rO)$ is the relative radial wave function. For $L=J\mp 1$, $L'=J'\mp 1$ and $S=1$ the tensor correlator, following Eq.~\eqref{tensorCtilde}, transforms the bra and the ket-state of the matrix element into a superposition of two states. This leads to a combination of four terms for the correlated matrix element 
\eqmulti{ \label{eq:corrme1}
 &\matrixe{n(J\mpS 1,1) J T}{\ccO_{r} \ccO_{\Omega}\, f(\rO) Y_{\JC\MC}(\hat{\rOV})\,  
   \cO_{\Omega}\cO_{r}}{n'(J'\mpS 1,1) J' T} = \\
  &= \int\!\dd{r}\,u_{n,J\mpS 1}^{\star}(r)\, u_{n',J'\mpS 1}(r)\;
     \corr{f}(r) \\[-2pt]
  &\quad\times\big[\matrixe{(J\mpS 1,1) J T}{Y_{\JC\MC}(\hat{\rOV})}{(J'\mpS 1,1) J' T}\, 
    \cos \corr{\theta}_{J}(r)  \cos \corr{\theta}_{J'}(r) \\ 
  &\quad+\;\, \matrixe{(J\pmS 1,1) J T}{Y_{\JC\MC}(\hat{\rOV})}{(J'\pmS 1,1) J' T}\, 
   \sin \corr{\theta}_{J}(r) \sin \corr{\theta}_{J'}(r)\\
  &\quad\pm\;\, \matrixe{(J\mpS 1,1) J T}{Y_{\JC\MC}(\hat{\rOV})}{(J'\pmS 1,1) J' T}\, 
    \cos\corr{\theta}_{J}(r) \sin\corr{\theta}_{J'}(r)\\
  &\quad\pm\;\, \matrixe{(J\mpS 1,1) J T}{Y_{\JC\MC}(\hat{\rOV})}{(J'\pmS 1,1) J' T}\, 
     \sin\corr{\theta}_{J}(r) \cos\corr{\theta}_{J'}(r) \big] \;.
}
An analogous expression is obtained for $L=J\mp 1$ and $L'=J'\pm 1$. For $L=J\pm 1$ and $L'=J'$, the tensor correlator affects only the bra-state and we get the simpler form
\eqmulti{ \label{eq:corrme3}
  &\matrixe{n(J\pmS 1,1) J T}{\ccO_{r} \ccO_{\Omega}\, f(\rO) Y_{\JC\MC}(\hat{\rOV})\,   
    \cO_{\Omega}\cO_{r}}{n'(J',1) J' T} = \\  
  &= \int\!\dd{r}\, u_{n,J\pmS 1}^{\star}(r)\, u_{n',J' 1}(r)\;
     \corr{f}(r) \\[-2pt]
  &\quad\times\big[ \matrixe{(J\pmS 1,1) J T}{Y_{\JC\MC}(\hat{\rOV})}{(J',1) J'T}\,
    \cos \corr{\theta}_{J}(r) \\
  &\quad\mp\;\, \matrixe{(J\mpS 1,1) J T}{Y_{\JC\MC}(\hat{\rOV})}{(J',1) J'T}\,
    \sin\corr{\theta}_{J}(r)\big] \;.
}

At this point we should like to remark, that the present UCOM-RPA scheme is self-consistent in two respects. First, the same correlated realistic NN interaction $\VO_{\UCOM}$ is used in the HF equations that determine the single-particle basis and in the RPA equations as residual interaction. Hence, the effective NN interaction which determines the ground-state properties, also determines the small amplitude motion around the nuclear ground state. This ensures that RPA amplitudes do not contain spurious components associated with the center-of-mass translational motion, which will be shown explicitly in Sec. \ref{secIII}. Second, the unitary transformation used to construct the effective interaction $\VO_{\UCOM}$ is also applied to the transition operators. Hence the effect of short-range central and tensor correlations is included consistently in all relevant observables.

%%%%%%%%%%%%%%%%%%%%%%%%%%%%%%%%%%%%%%%%%%%%%%%%%%%%%%%%%%%%%%%%%%%%%%%%%%%%%%%
\section{\label{secIII}Tests of the UCOM-RPA implementation}

In order to ensure that the UCOM-RPA scheme is properly implemented, and to probe its self-consistency, we perform several stringent tests. In particular, in studies of the multipole response of closed-shell nuclei across the nuclear chart, the following conditions need to be fulfilled: (a) the spurious excitation corresponding to a translation of a nucleus decouples as a zero-energy excitation mode,
(b) the transition strength should exhaust the sum rules, and (c) the excitation energies of giant resonances and low-lying states converge to stable solutions independent of the energy cut-off parameter. As will be demonstrated in this section, the present UCOM-RPA scheme is fully consistent with these conditions.

%%%%%%%%%%%%%%%%%%%%%%%%%%%%%%%%%%%%%%%%%%%%%%%%%%%%%%%%%%%%%%%%%%%%%%%%%%

\subsection{\label{ssecIII-1}Spurious solutions of the UCOM-RPA equations}

%%%%%%%%%%%%%%%%%%%%%%%%%%%%%%%%%%%%%%%%%%%%%%%%%%%%%%%%%%%%%%%%%%%%%%%%%%

Whenever the generator of a continuous symmetry for a general 
two-body Hamiltonian
does not commute with the original single-particle density, 
it produces a spurious zero-energy solution of the RPA 
equations~\cite{Rin.80}. 
It corresponds to a mode which is not related to an intrinsic
excitation of the system, but in fact to a collective motion 
without restoring force. In our particular case, a spurious dipole 
mode is associated with translation, i.e. the center-of-mass motion. 
Ideally, if the RPA scheme is built on the self-consistent wave
functions and single-particle energies, the spurious excitation 
should decouple from the physical states at exactly zero 
energy. In practical calculations, however, due to the truncation of
the $ph$ configuration space and
inconsistencies between the ground state and RPA equations, the spurious
state is separated at energies larger than zero. In this case the 
physical states may be more or less mixed with the spurious
response, leading to seriously overestimated strength distributions,
especially in the low-energy region~\cite{Col.00}.
Violations of self-consistency in the Hartree-Fock based RPA
may also cause a spurious enhancement of the isoscalar monopole mode energy
for spin unsaturated systems~\cite{Agr.04}.

\begin{figure}
\includegraphics[width=0.9\columnwidth]{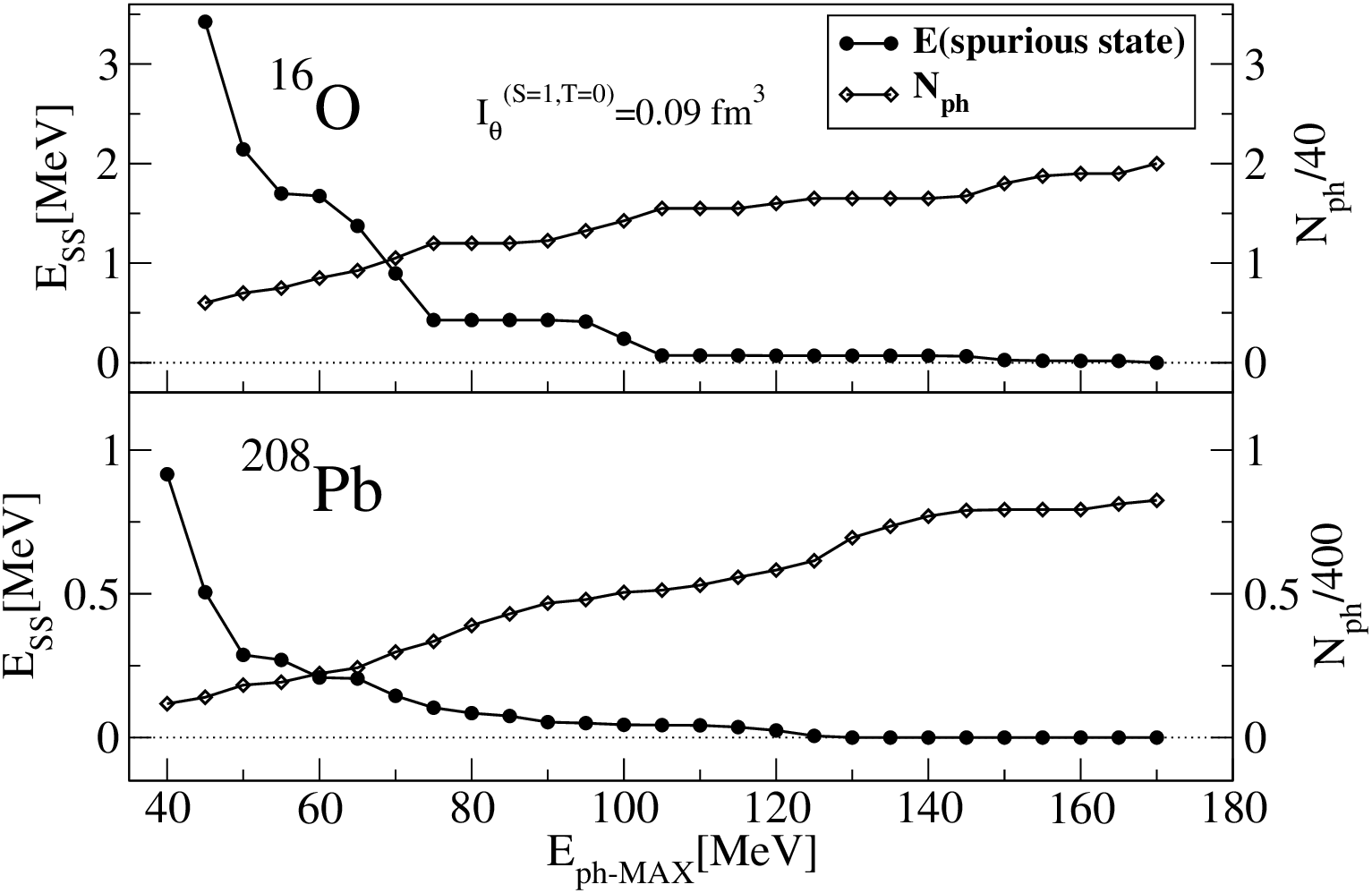}
\caption{The number of $ph$ configurations in the UCOM-RPA (right scale)
and the energy of the spurious center-of-mass state (left scale) as a function of the cut-off for the maximal energy of the $ph$ transitions. Two sample nuclei
are considered, $^{16}$O and $^{208}$Pb. The correlated Argonne
V18 interaction is employed, using the standard correlators~(Sec.~\ref{ssecII-1}).
} 
\label{figspur1}
\end{figure}

The truncation of the actual RPA configuration space may strongly affect
the separation of spurious states. The size of the $ph$ space is determined
by a cut-off parameter $E_{ph-MAX}$ which corresponds to the
maximal allowed energy for $ph$ excitations. In Fig.~\ref{figspur1} 
we display the energy of the spurious state $E_{SS}$ as a function of 
the energy cut-off $E_{ph-MAX}$, for $^{16}$O and $^{208}$Pb.
The isoscalar dipole operator is employed,
\begin{equation}
\QC^{\TC=0}_{1\MC} = e\sum_{i=1}^{A} \xO_i^3 Y_{1\MC}(\hat{\xOV}_i), 
\label{isgdrtroperator}
\end{equation}
which leads to both, $1 \hbar \omega $ and $3 \hbar \omega $ $ph$
excitations.
In the same graphs, we also present the corresponding number of $ph$
configurations $N_{ph}$ which enter into the RPA equations. We employ
the correlated Argonne V18 interaction with the standard correlators,
as discussed in Sec.~\ref{ssecII-1}.
As the number of $ph$ configurations increases, the spurious
state converges towards zero excitation energy for both nuclei. This necessitates relatively
large spaces with 80 and  330 configurations
for $^{16}$O and $^{208}$Pb, respectively. The accuracy of our method
is exemplified by the energies of the spurious states : 0.01 MeV 
for $^{16}$O, and 0.05 MeV for $^{208}$Pb, which appear to be 
lower than in recently developed fully self-consistent (Q)RPA 
models based on phenomenological nuclear interactions~\cite{Paa.03,Ter.04}.

\begin{figure}
\includegraphics[width=0.9\columnwidth]{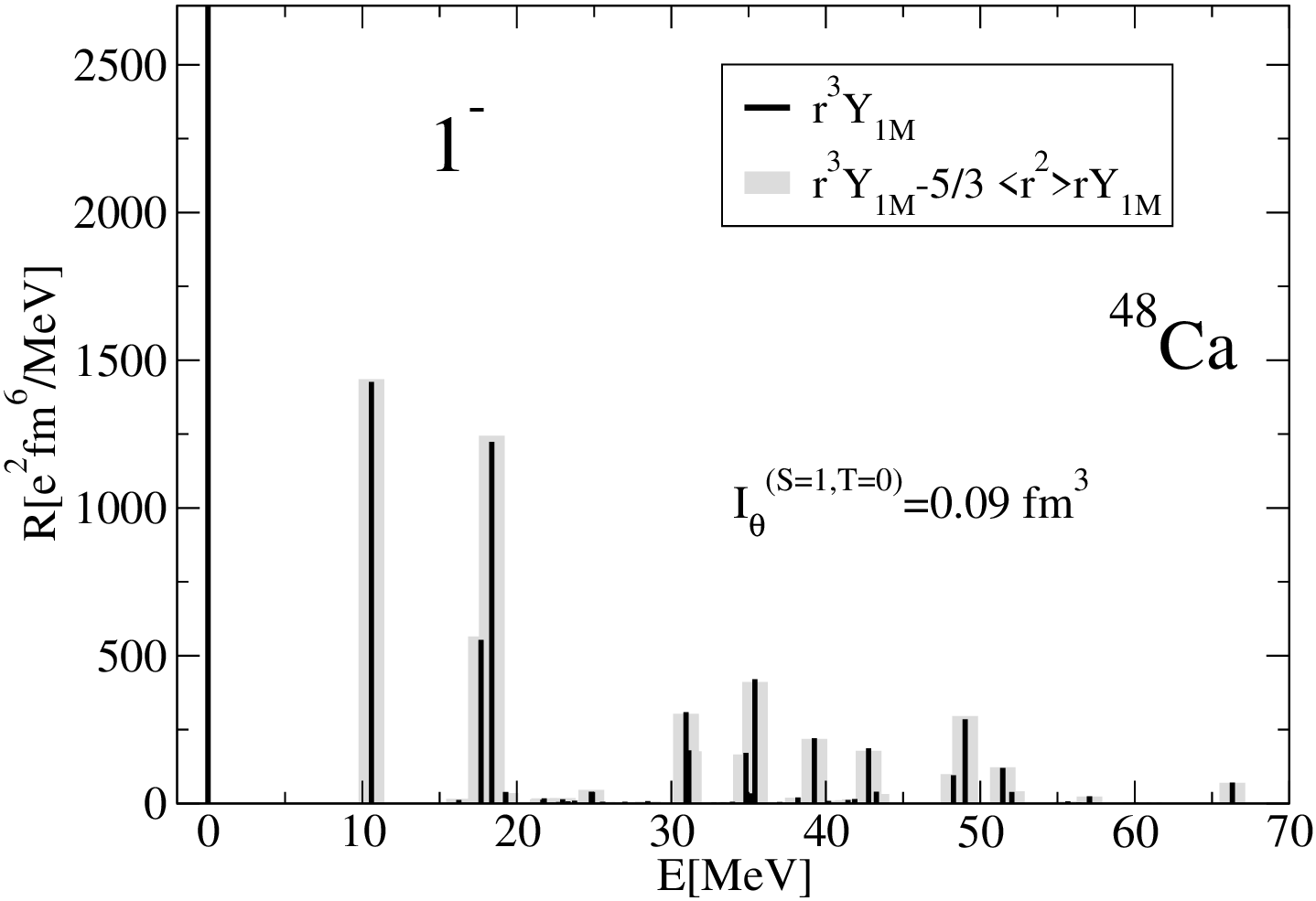}
\vspace{1cm}
\caption{The UCOM-RPA isoscalar dipole transition spectra for $^{48}$Ca, by employing
the isoscalar dipole operator without (thin solid line) and with 
(wide grey line) the correction term for the spurious center-of-mass motion 
(Argonne V18, $I_{\vartheta}^{(S=1,T=0)}$=0.09 fm$^3$).
}
\label{figspur2}
\end{figure}

Next, we employ effective dipole transition operators, 
\begin{equation}
\QC^{\TC=0}_{1\MC} = e\sum^{A}_{i=1} \, \left( \xO_i^3 - \frac{5}{3} \, 
\langle  \xO^2 \rangle_{_0} \xO_i \right) Y_{1\MC}(\hat{\xOV}_i),
\label{isooperator}
\end{equation}
%
%ISOVECTOR DIPOLE OPERATOR  
\begin{equation}
\QC_{1\MC}^{\TC=1} = e\sum_{i=1}^A \left( \tau_z^{(i)}-\frac{N-Z}{2A}
\right)\xO_{i} Y_{1\MC}(\hat{\xOV}_i), 
\label{ivdipole1}
\end{equation} 
which explicitly contain the center-of-mass correction terms on the
operator level~\cite{Giai.81}. The high accuracy of the UCOM-RPA method
is illustrated in Fig.~\ref{figspur2} for $^{48}$Ca,
where we compare the isoscalar dipole transition strength distribution for the
uncorrected operator ~(\ref{isgdrtroperator}), and the operator
with the center of mass correction term~(\ref{isooperator}).
The two spectra are almost identical, and the major
difference is only in the strong
transition of the spurious state at 0.005 MeV, obtained for the
operator without a center
of mass correction. When the correction term is included, the spurious state 
is completely removed, whereas transitions corresponding to actual dipole
vibrations are practically unaffected. This means that the physical excited
states are  free of spurious contaminations, owing to the full
self-consistency of our method. In  Fig.~\ref{figspur2} one can also observe a double hump 
structure that is characteristic for the isoscalar giant dipole resonance (ISGDR).
It is composed of a high-energy part corresponding to a compression
mode, and a low-energy part which might be of different nature~\cite{Vre.02}.

%%%%%%%%%%%%%%%%%%%%%%%%%%%%%%%%%%%%%%%%%%%%%%%%%%%%%%%%%%%%%%%%%%%%%
\subsection{\label{ssecIII-2}Sum rules of the multipole strength in
the UCOM-RPA scheme}
%%%%%%%%%%%%%%%%%%%%%%%%%%%%%%%%%%%%%%%%%%%%%%%%%%%%%%%%%%%%%%%%%%%%%

The sum rules of various collective excitation modes is of
particular interest, since their values represent a useful
test for the RPA~\cite{Rin.80}. In the present study,
we examine the energy-weighted sum rule (EWSR)~\cite{Ter.04}
for the isoscalar monopole, 
\begin{equation}
S(E0) = \frac{2 \hbar ^2 e^2}{m} \left( N \left< \xOV_n^2 \right> + 
Z \left< \xOV_p^2 \right> \right),
\label{sre0}
\end{equation}
and isoscalar quadrupole excitations,
\begin{equation}
S(E2) = \frac{25 \hbar ^2 e^2}{4\pi m} \left( N \left< \xOV_n^2 \right> + 
Z \left< \xOV_p^2 \right> \right).
\label{sre2}
\end{equation}

\begin{figure}
\includegraphics[width=0.9\columnwidth]{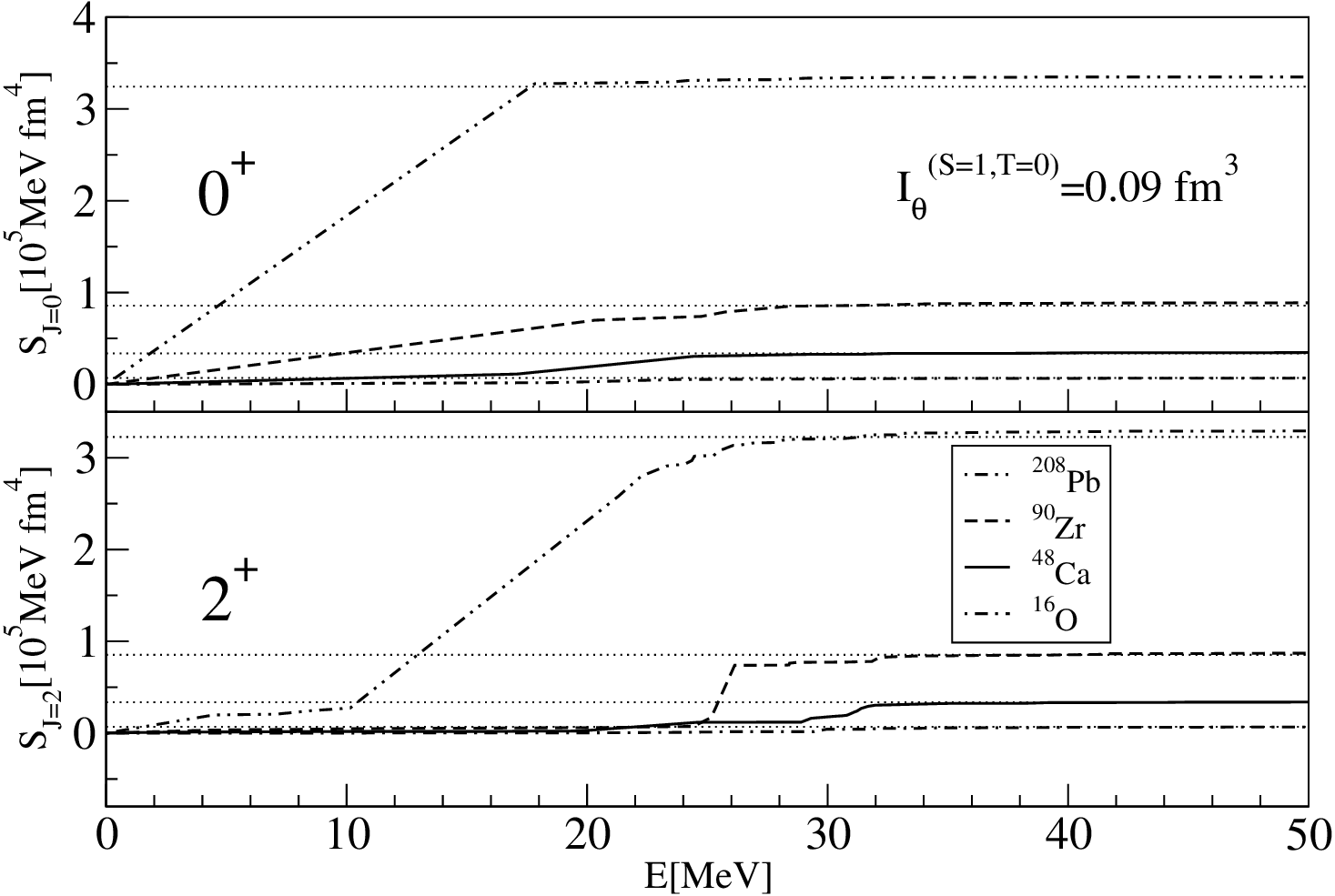}
\vspace{1cm}
\caption{The cumulative sum of the UCOM-RPA energy weighted strength for
the isoscalar monopole (upper panel) and the quadrupole response (lower panel), in comparison
with the energy weighted sum rules (horizontal dotted lines)~(\protect\ref{sre0}),(\protect\ref{sre2}), for
$^{16}$O, $^{48}$Ca, $^{90}$Zr, and $^{208}$Pb
(Argonne V18, $I_{\vartheta}^{(S=1,T=0)}$=0.09 fm$^3$).}
\label{figsumrule}
\end{figure}

In Fig.~\ref{figsumrule}, we plot the cumulative energy weighted sum of the 
transition strength for given excitation energy $E<50$ MeV in the
closed-shell nuclei $^{16}$O, $^{48}$Ca, $^{90}$Zr, and $^{208}$Pb.
The upper and lower panels display the summed energy weighted strength for
the isoscalar monopole and quadrupole cases, respectively. One can observe that
beyond 40 MeV the sums converge to their final values, which are in good agreement with
the EWSR from Eqs.~(\ref{sre0}) and~(\ref{sre2}). 
We have also confirmed that UCOM-RPA essentially exhausts the EWSR
with maximal discrepancies of $\pm 3\%$ in other closed-shell nuclei. 
Accordingly, we conclude that the completeness properties and consistency
are accurately fulfilled in the UCOM-RPA approach.

In the following, we will average the discrete UCOM-RPA strength
distributions with a Lorentzian function, and therefore obtain
the continuous strength function,
\begin{equation}
R^{\TC}_{\JC}(E) = \sum_{\nu}~B^{\TC}_{\JC}(\omega_{\nu})~\frac{1}{\pi}
~\frac{\Gamma/2}
{ (E - \omega_{\nu} )^2 +(\Gamma/2)^2}.
\label{Lorentz}
\end{equation}
The width of the Lorentzian distribution is fixed to the 
arbitrary value $\Gamma$=2 MeV. The Lorentzian function
~(\ref{Lorentz}) is defined in a way to fulfill the condition that the 
sum of the energy weighted response is equal for the discrete distribution
and the continuous strength function, 
\begin{equation}
S^{\TC}_{\JC} = \sum_{\nu}E_{\nu}  B^{\TC}_{\JC}(\omega_{\nu})
 = \int \dd E\, E R^{\TC}_{\JC}(E).
\label{ewsr}
\end{equation}
%
%%%%%%%%%%%%%%%%%%%%%%%%%%%%%%%%%%%%%%%%%%%%%%%%%%%%%%%%%%%%%%%%%%%%%%%%%%%%%%%%%%%%%
\subsection{\label{secIV-4}The role of correlated multipole transition operators}
The UCOM-RPA observables describing collective excitation phenomena are evaluated
by a consistent application of the same unitary transformation as for
the nuclear Hamiltonian. In this section, we test the relevance of the correlated
transition operators for the multipole strength distributions.

\begin{figure}
\includegraphics[width=0.9\columnwidth]{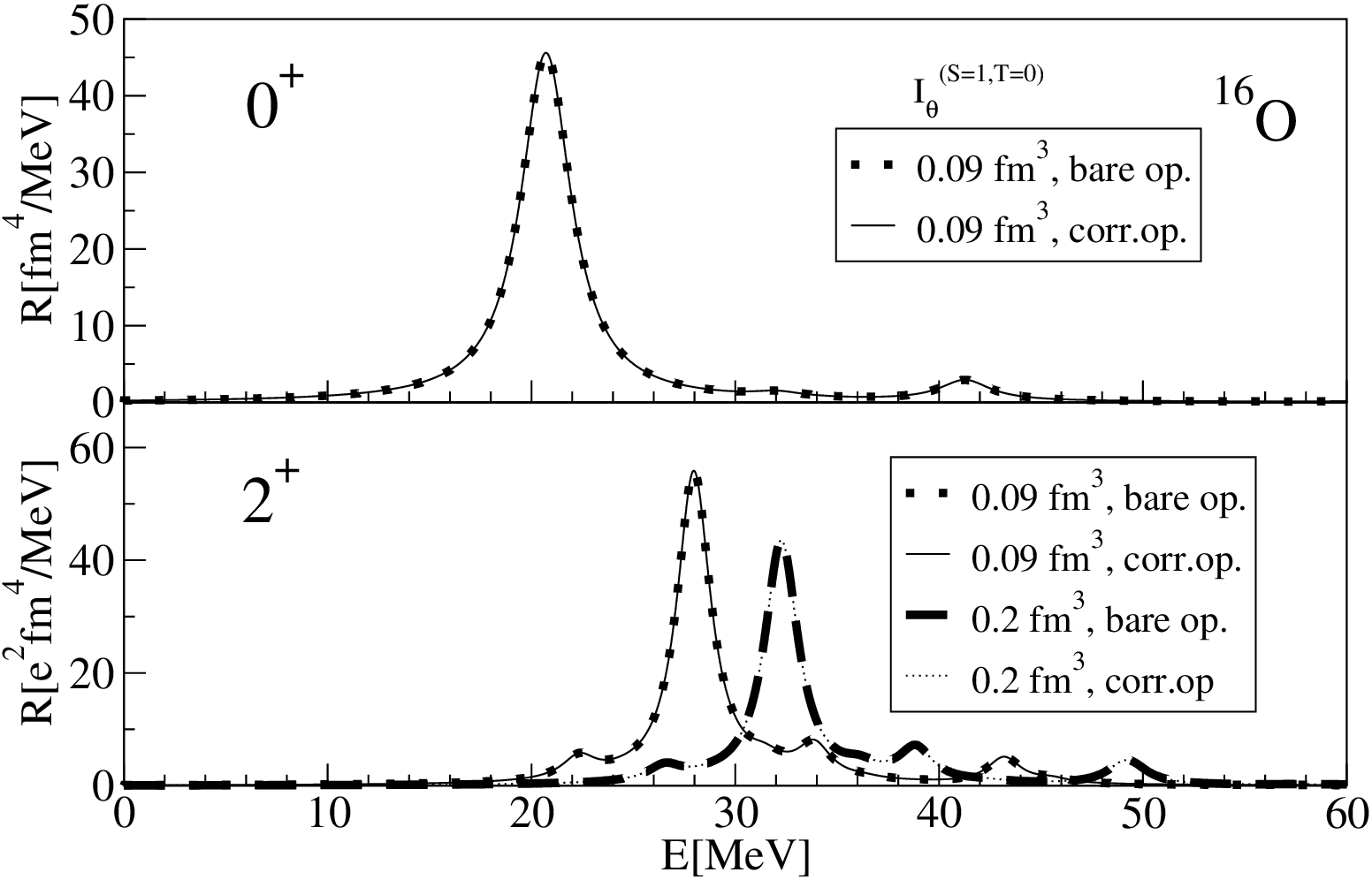}
\vspace{1cm}
\caption{The isoscalar monopole (upper panel) and quadrupole (lower panel)
strength distributions
for $^{16}$O, calculated with the bare and correlated transition operators.
The correlated Argonne V18 interaction is used, with the constraints $I_{\vartheta}^{(S=1,T=0)}$=0.09 and 0.2 fm$^3$ for the range of the
tensor correlator ($N_{max}= 8$ for 
the harmonic oscillator basis).}
\label{figoper1}
\end{figure}

\begin{figure}
\includegraphics[width=0.9\columnwidth]{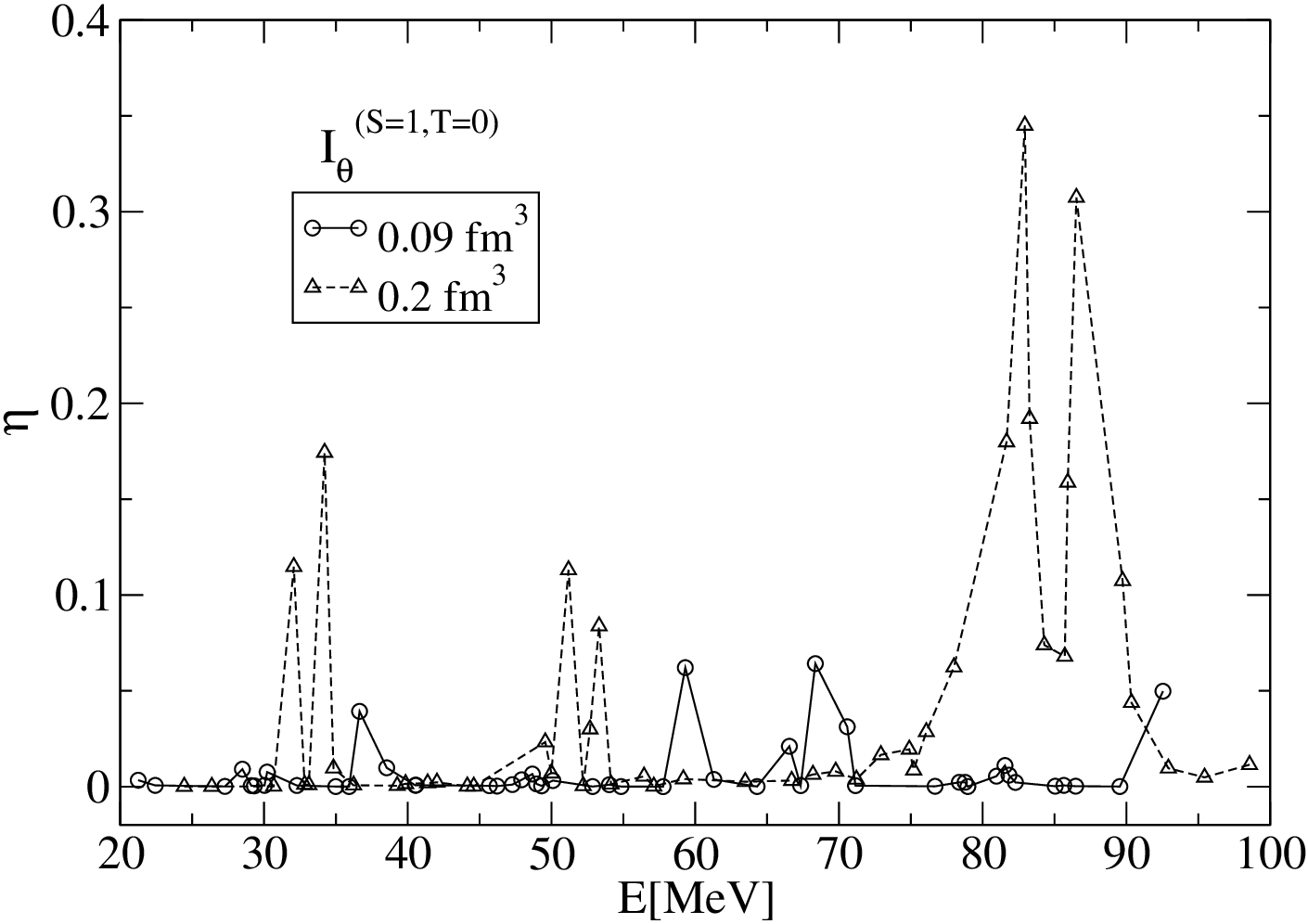}
\vspace{1cm}
\caption{The absolute value of the ratio of the transition matrix elements
for the two-body terms of correlated operator against those of the bare operator.
The tensor correlators are constrained by $I_{\vartheta}^{(S=1,T=0)}$=0.09 fm$^3$
and $I_{\vartheta}^{(S=1,T=0)}$=0.2 fm$^3$ (Argonne V18, $N_{max}= 8$).}
\label{figoper2}
\end{figure}

In Fig.~\ref{figoper1}, the isoscalar monopole and quadrupole responses
in $^{16}$O are
displayed for the two cases: (i) with the bare
multipole operators 
(\ref{monoperator}),~(\ref{istroperator}),~(\ref{tran1body}), and  
(ii) with the correlated multipole operators constructed
by the unitary transformation, Eq.~(\ref{strength}). The two transition
strength distributions are essentially identical. 
In order to probe the relevance of the tensor correlations
for the quadrupole response (in the monopole case, only the central
correlations are active), we construct a new tensor correlator
with a long range, constrained by 
$I_{\vartheta}^{(S=1,T=0)}$=0.2 $\text{fm}^3$. However, even in this extreme
case, the correlated quadrupole response resembles the one obtained with 
the bare operator (Fig.~\ref{figoper1}). In order to understand quantitatively the contributions of the
correlated quadrupole operators with various ranges of the tensor correlation
functions, in Fig.~\ref{figoper2} we plot the ratio
of the transition matrix elements for the two-body term of the correlated
operator against the one with the bare operator for each RPA eigenvalue, i.e.
\begin{equation}  
\eta =  \bigg|
\frac{\langle \nu || \corr{\QC}^{[2]}|| 0 \rangle }{\langle \nu || \QC || 0 \rangle}\bigg|.
\label{etapar}  
\end{equation} 
For the tensor correlator
with the constraint $I_{\vartheta}^{(S=1,T=0)}$=0.09 $\text{fm}^3$, corrections to the
bare operator are rather small. On the other hand, in the case of tensor
correlator with longer range ($I_{\vartheta}^{(S=1,T=0)}$=0.2 $\text{fm}^3$),
we notice more pronounced relative contributions from the two-body terms. However,
their small absolute values result in negligible corrections of the overall transition spectra.
It is interesting to note,
that this observation is in agreement with the study of effective operators in the no-core shell
model within the $2\hbar\Omega$ model space, where the B(E2) values
are very similar for the bare and the effective operator which includes
the two-body contributions~\cite{Ste.05}.

Due to the high computational effort involved in evaluating the non-diagonal two-body matrix elements
for the multipole operators, we used a smaller oscillator
basis here ($N_{max}= 8$). We have verified that the effect
of two-body terms of the correlated operators on the
transition spectra is very small, regardless of the size of the basis.

%=========================================================================

%%%%%%%%%%%%%%%%%%%%%%%%%%%%%%%%%%%%%%%%%%%%%
% Section IV
\section{\label{secIV}Multipole excitations in the UCOM-RPA framework}
%%%%%%%%%%%%%%%%%%%%%%%%%%%%%%%%%%%%%%%%%%%%%

Among collective modes of
excitations, giant resonances have been a very active topic of 
nuclear physics in the past few decades, both 
theoretically~\cite{Row.70,BT.75,Spe.81,Dum.83,Daw.90,Ham.97,Col.00,Rod.02}
and experimentally~\cite{Ber.81,Gar.04}.
Collective modes are of particular importance for 
theoretical models, because their underlying dynamics provide
direct information about different effective nuclear interactions
and methods employed to solve the nuclear many-body problem. 
In the following, using the UCOM-RPA scheme, we evaluate the 
excitation energies and transition strengths of giant multipole
resonances, and probe their sensitivity on the properties of
the correlation functions. 

\subsection{\label{secIV-1}Giant monopole resonances}

The isoscalar giant monopole resonance (ISGMR) is a spherically
symmetric oscillation or compression of the nucleus, i.e. a breathing
mode where neutrons and protons move in phase. 
The ISGMR excitation energy is related to the compressibility
of nuclear matter $K_{nm}$,
which defines basic properties of nuclei, supernovae explosions,
neutron stars, and heavy ion collisions~\cite{Bla.80}.
Therefore, it is rather important to provide information
about ISGMR from various models based on different effective
interactions.

\begin{figure}
\includegraphics[width=0.9\columnwidth]{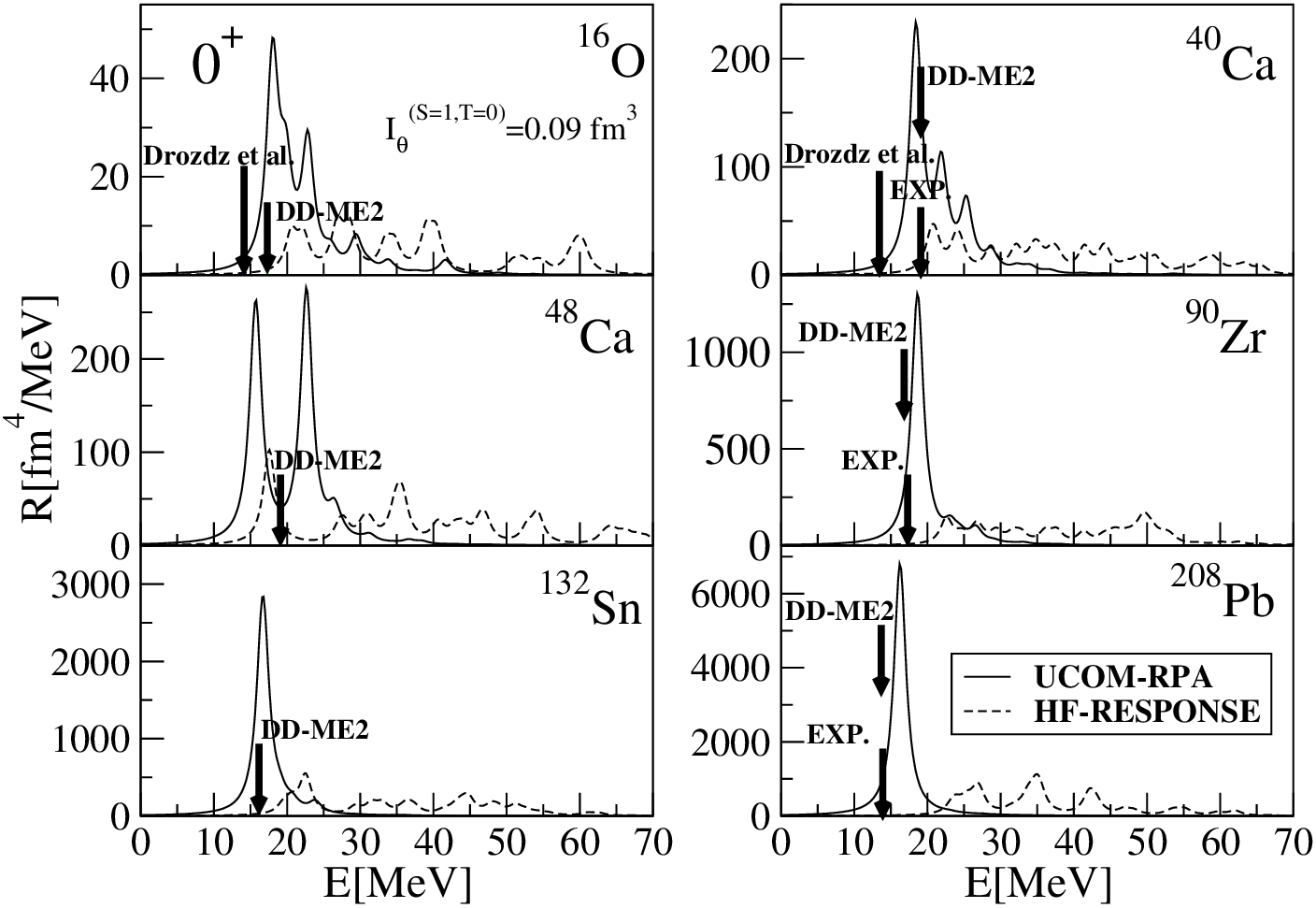}
\vspace{1cm}
\caption{The UCOM-RPA transition strength distributions of the ISGMR (solid line), in comparison with the unperturbed HF spectra (dashed line).
The correlated Argonne V18 interaction is used, with the standard 
constraint ($I_{\vartheta}^{(S=1,T=0)}$=0.09 fm$^3$) on the range of the tensor correlators.
The experimental
data~\protect\cite{You.99,Shl.93,Sha.88,You.04}, results
from the nonrelativistic (Dro{\. z}d{\. z} et al.)~\protect\cite{Dro.90},
and the relativistic RPA (DD-ME2)~\cite{NVR.02,LNVR.05} are denoted by arrows.}
\label{figmono1}
\end{figure}

The UCOM-RPA calculated ISGMR strength distributions are displayed
in Fig.~\ref{figmono1} for a series of closed-shell nuclei from
$^{16}$O to $^{208}$Pb. Both the HF single-nucleon basis and
the RPA transition strength are calculated with
the correlated Argonne V18 interaction, using the standard
ranges of the central and tensor correlation functions.
For comparison, the unperturbed HF response in the $0^{+}$ channel is also shown. 
One can notice that the unperturbed spectra are widely spread in the
energy region $\approx$20-70 MeV, as a direct consequence
of the relatively low 
level density of the UCOM-HF single-particle spectra (Figs.~\ref{figspectra1}
and \ref{figspectra2}). However, when we include the RPA 
residual interaction, which is attractive in the isoscalar
channel, most of the unperturbed strength is pushed to lower energies,
generating the collective ISGMR mode. For lighter nuclei $^{16}$O, $^{40}$Ca,
and $^{48}$Ca, the ISGMR is fragmented into two or three peaks,
whereas for $^{90}$Zr, $^{132}$Sn, and $^{208}$Pb the 
ISGMR mode is strongly collective, resulting essentially in a single
peak. The calculated ISGMR strength distributions shown in Fig.~\ref{figmono1}
are also compared with the available experimental data from
$(\alpha,\alpha)$ \cite{You.99,Shl.93,You.04} and ($^3$He,$^3$He) 
scattering \cite{Sha.88}, and with results from the nonrelativistic RPA~\cite{Dro.90} and
relativistic RPA~\cite{NVR.02} based on a new interaction with density-dependent
meson-nucleon couplings (DD-ME2)~\cite{LNVR.05}.
The UCOM-RPA excitation energies are in general very close to the values
from the other studies. In nuclei where the breathing mode is well 
established ( $^{90}$Zr, $^{208}$Pb), the centroid energies for the
standard correlator are slightly overestimated.
The small discrepancies ($\approx$1-3 MeV) obtained for the standard
correlator could originate
from the missing long-range correlations beyond
the simple mean-field level, the missing genuine three-body
interaction, and the two-body approximation in the UCOM method,
as discussed in Sec.~\ref{ssecII-1}. 

\begin{figure}
\includegraphics[width=0.9\columnwidth]{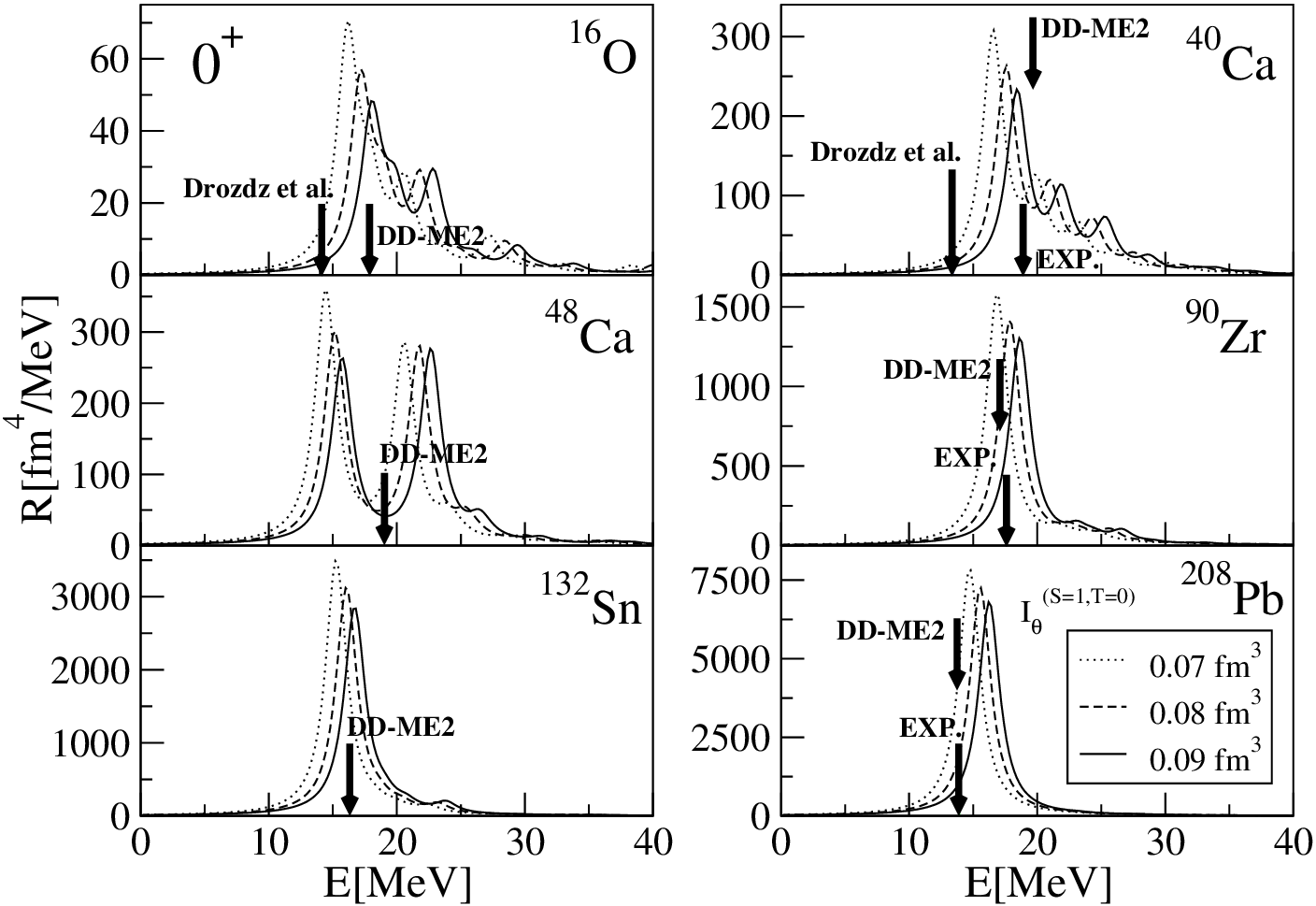}
\vspace{1cm}
\caption{The calculated UCOM-RPA strength distributions of the ISGMR (Argonne V18),
using different restrictions on the range of the tensor correlator ($I_{\vartheta}^{(S=1,T=0)}$=0.07, 0.08, and 0.09 fm$^3$). Results from the other studies, denoted by arrows, are the same as in Fig.~\protect\ref{figmono1}.}
\label{figmono2}
\end{figure}

Next, we address the question to which extent the UCOM-RPA transition
spectra are sensitive to the ranges of the correlators used in the unitary transformation to construct the correlated interaction. 
In Fig.~\ref{figmono2}, the calculated ISGMR strength distributions are displayed
for the correlated Argonne V18 interaction with different constraints on the range
of the tensor correlator, $I_{\vartheta}^{(S=1,T=0)}$=0.07, 0.08, and 0.09 fm$^3$.
One can observe that decreasing of this range systematically pushes the
transition strength towards lower energies. In particular, by decreasing the 
range of the tensor correlator, i.e. its constraint from $I_{\vartheta}^{(S=1,T=0)}$=0.09 fm$^3$
towards 0.07 fm$^3$, the excitation energy of ISGMR lowers
by $\approx$4 MeV. By lowering the range of the
tensor correlation functions, the density of 
single-particle spectra increases and the agreement
of the ISGMR excitation energies with experimental data is improved.

%%%%%%%%%%%%%%%%%%%%%%%%%%%%%%%%%%%%%%%%%%%%%%%%%%%%%%%%%%%%%%%%%%%%%%%%%%%%%%
\subsection{\label{secIV-2}Giant dipole resonances}

\begin{figure}
\includegraphics[width=0.9\columnwidth]{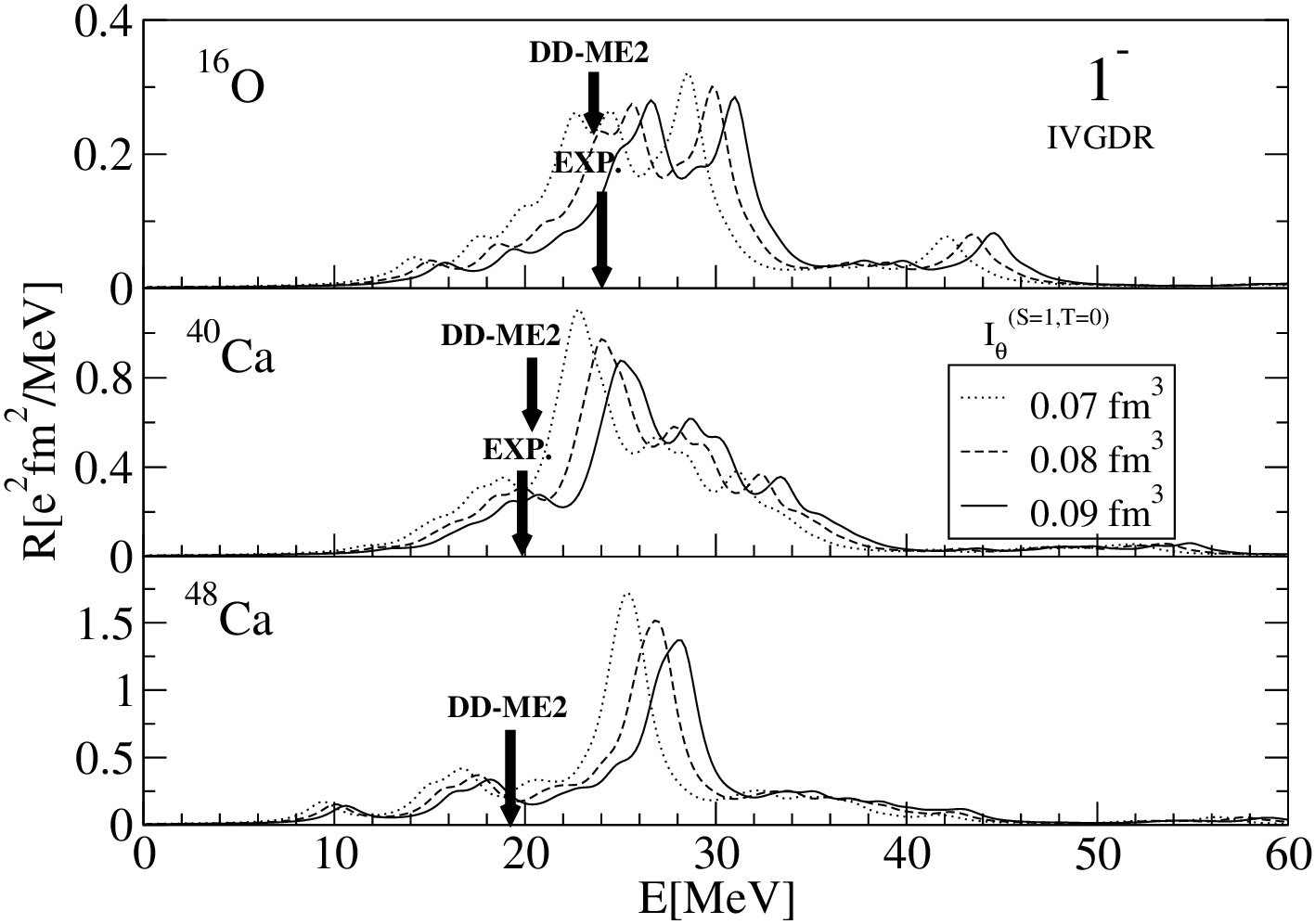}
\vspace{1cm}
\caption{The UCOM-RPA strength distributions for the IVGDR in $^{16}$O, $^{40}$Ca, and $^{48}$Ca. The calculations are based on the correlated Argonne V18 interaction,
using different constraints on the tensor correlator range ($I_{\vartheta}^{(S=1,T=0)}$= 0.07, 0.08, and 0.09 fm$^3$). The experimental~\protect\cite{Ber.75} and theoretical IVGDR energies from the relativistic RPA (DD-ME2)~\protect\cite{NVR.02,LNVR.05} are denoted by arrows.
}
\label{figdip1}
\end{figure}

\begin{figure}
\includegraphics[width=0.9\columnwidth]{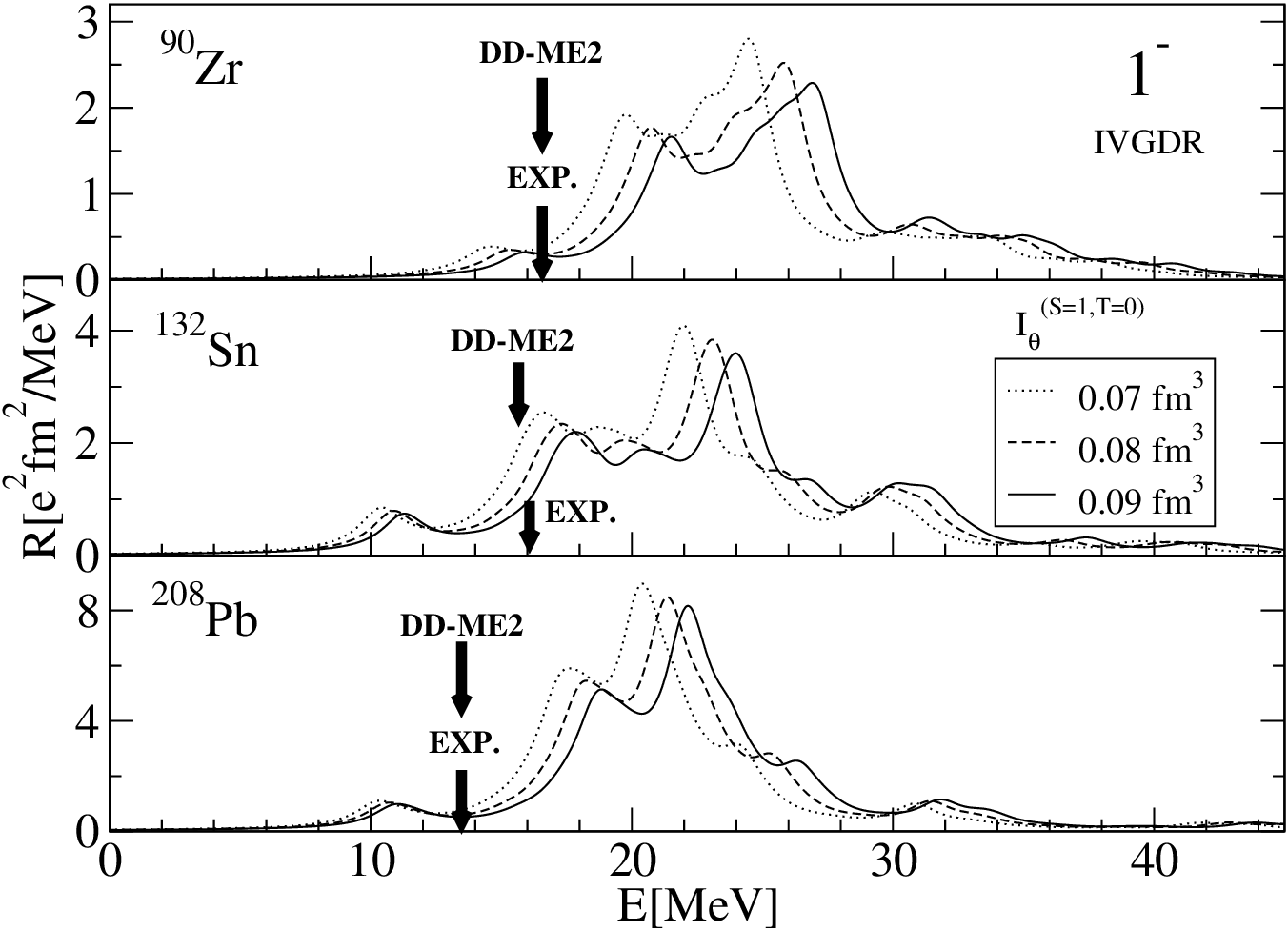}
\vspace{1cm}
\caption{The same as Fig.~\ref{figdip1}, but for $^{90}$Zr,
$^{132}$Sn, and $^{208}$Pb. The experimental data~\protect\cite{Adr.05,Ber.75,Poe.89,Rit.93} and
the relativistic RPA (DD-ME2) energies~\protect\cite{NVR.02,LNVR.05} are shown by arrows.}
\label{figdip2}
\end{figure}

The isovector giant dipole resonances (IVGDR) have recently been studied
very extensively, in parallel with the renewed interest
in the low-lying dipole strength in nuclei away from the
valley of stability~\cite{Sag.99,Lei.01,Ma.04,Paa.05,Gor.04,Adr.05,Zil.05}.
Here we employ the UCOM-RPA to evaluate the IVGDR strength
distributions in light nuclei $^{16}$O, $^{40}$Ca, and $^{48}$Ca,
using the correlated Argonne V18 interaction with
different constraints on the ranges of the tensor part of the correlator,
$I_{\vartheta}^{(S=1,T=0)}$=0.07, 0.08, and 0.09 fm$^3$ (Fig.~\ref{figdip1}).
In lighter nuclear systems, UCOM-RPA provides a collective
character of IVGDR, which is distributed over several dominant peaks.
We have also extended our study of IVGDR to heavier nuclear systems
$^{90}$Zr, $^{132}$Sn, and $^{208}$Pb (Fig.~\ref{figdip2}). 
The calculated dipole response is compared with 
experimental data~\cite{Adr.05,Ber.75,Poe.89,Rit.93} and with
the theoretical excitation energies from the relativistic
RPA~\cite{NVR.02} based on DD-ME2 interaction~\cite{LNVR.05}.
In all nuclei under consideration, the resulting IVGDR
strength distributions
display rather wide resonance-like structures. 
The decrease in the range of the tensor correlator, 
i.e. its constraint $I_{\vartheta}^{(S=1,T=0)}$=0.09 fm$^3$
towards 0.07 fm$^3$, results in lower IVGDR peak energies
by $\approx$2-3 MeV.

The calculated IVGDR strength distributions
systematically result in higher excitation
energies than the values from other studies.
For the short-range correlator ($I_{\vartheta}^{(S=1,T=0)}$=0.07 fm$^3$),
the calculated transition strength
appears to be in fair agreement with experimental data only for $^{16}$O. 
However, for other nuclear systems, in comparison with experimental data
and other theoretical studies,
the UCOM-RPA overestimates the IVGDR centroid energies by
$\approx$3-7 MeV. This difference can serve as a
direct measure of the missing correlations
and three-body contributions in the UCOM-RPA scheme.
Inclusion of the three-body interaction
and long-range correlations beyond the simple RPA method,
would probably to a large extent resolve this discrepancies with the other studies. 
Therefore, the $1^-$ channel is particularly convenient to probe 
the effects of the missing correlations and
three-body contributions.

%%%%%%%%%%%%%%%%%%%%%%%%%%%%%%%%%%%%%%%%%%%%%%%%%%%%%%%%%%%%%%%%%%%%%%%
\subsection{\label{secIV-3}Giant quadrupole resonances}

Giant quadrupole resonances are comprised of 2$\hbar\omega$ 
$ph$ configurations coupled by the residual interaction \cite{Ber.81}. 
In addition to the resonance-like structure corresponding to the isoscalar
giant quadrupole resonance (ISGQR), in the isoscalar channel
the interaction also generates pronounced
0$\hbar\omega$ low-lying quadrupole states, which have been a subject
of various recent theoretical studies~\cite{Ter.02,Mat.01,Kha.00,Kha.02,Yam.04}.
The low-lying $2^+$ states provide valuable information about the
properties of the effective interaction~\cite{Fle.04,Paaqu.05}.
 
\begin{figure}
\includegraphics[width=0.9\columnwidth]{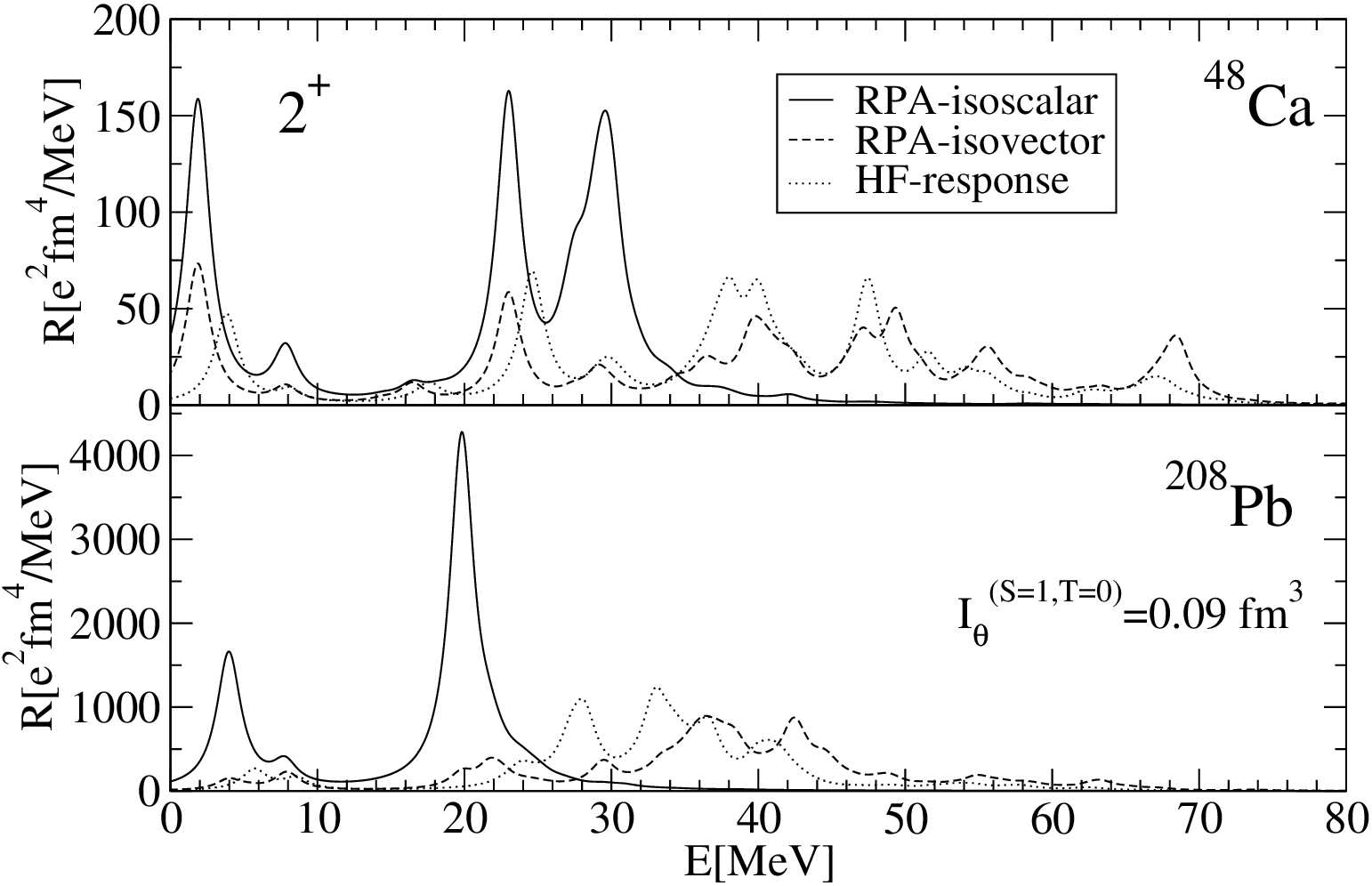}
\vspace{1cm}
\caption{The ISGQR and IVGQR strength distributions for $^{48}$Ca and
$^{208}$Pb, in comparison with the unperturbed HF response
(Argonne V18, $I_{\vartheta}^{(S=1,T=0)}$=0.09 fm$^3$).}
\label{figquad1}
\end{figure}

The isoscalar and isovector quadrupole transition strength distributions
are displayed in Fig.~\ref{figquad1} for representative cases of medium-mass
($^{48}$Ca)
and heavy nuclei ($^{208}$Pb). The UCOM-RPA calculations are based on
the correlated Argonne V18 interaction with
the standard ranges of central and tensor correlation functions. 
The unperturbed HF response is also shown, representing a broad distribution
of the strength in the region $\approx$2-90 MeV which
directly reflects the properties of the UCOM-HF single
particle-spectra (Sec. \ref{ssecII-2}). When the residual
interaction is included in RPA, the correlated realistic NN potential
generates pronounced low-lying quadrupole excitations in the isoscalar
channel, and strongly collective states at higher energies
corresponding to the ISGQR. For $^{48}$Ca,
the ISGQR is fragmented into two main peaks in the region $\approx$22-30 MeV,
whereas for the case of $^{208}$Pb, a single highly collective peak is generated
at 20.1 MeV. In comparison to the unperturbed spectra, the transition strength in the 
isoscalar channel is systematically pushed towards lower energies. On the other
hand, in the isovector channel which is repulsive, 
the strength distributions are in general moved towards the
energies above the unperturbed response, resulting in broad
structures of the transition strength. 
  
\begin{figure}
\includegraphics[width=0.9\columnwidth]{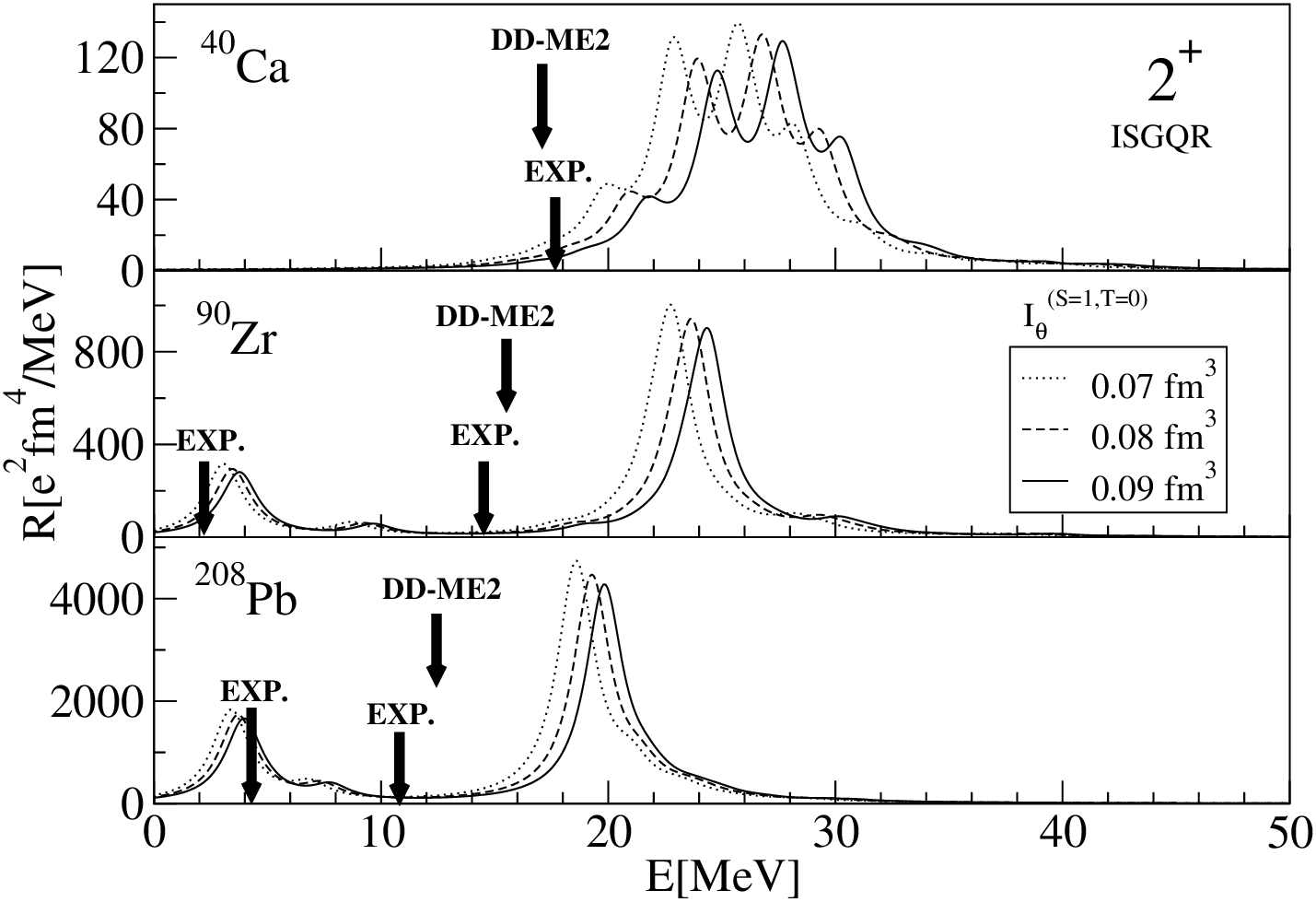}
\vspace{1cm}
\caption{The UCOM-RPA strength distributions for the ISGQR in $^{40}$Ca, $^{90}$Zr, and $^{208}$Pb.
The correlated Argonne V18 interaction is used, with different ranges of the tensor correlator ($I_{\vartheta}^{(S=1,T=0)}$=0.07, 0.08, and 0.09 fm$^3$).
The experimental ISGQR excitation energies~\protect\cite{Ber.79}, the low-lying states~\protect\cite{Tik.01,Hei.82}, and
the relativistic RPA energies~\protect\cite{NVR.02,LNVR.05} are denoted by arrows.}
\label{figquad2}
\end{figure}

In Fig.~\ref{figquad2}, we show the UCOM-RPA isoscalar quadrupole
transition strength distributions for $^{40}$Ca, $^{90}$Zr, and $^{208}$Pb.
The correlated Argonne V18 interaction is employed, with various 
constraints on the range of the tensor correlator, 
$I_{\vartheta}^{(S=1,T=0)}$=0.07, 0.08, and 0.09 fm$^3$. 
The calculated ISGQR strength distributions are also compared with the
experimental data \cite{Ber.79}, and the relativistic RPA calculations~\cite{NVR.02}
with density-dependent DD-ME2 interaction~\cite{LNVR.05}.
In all cases, the residual interaction constructed from the correlated realistic NN
interaction is attractive in the isoscalar channel, generating strongly
collective peaks corresponding to ISGQR. In addition, in the case of
$^{90}$Zr, and $^{208}$Pb, UCOM-RPA also results in the pronounced
low-lying quadrupole states. 
The energy of the low-lying quadrupole state in $^{90}$Zr is slightly 
higher than the experimental value (2.18~MeV~\cite{Tik.01}), and for
$^{208}$Pb it is rather well described (4.08~MeV~\cite{Hei.82}). 
However, the correlated realistic NN interaction
is not sufficient for a quantitative description of the ISGQR excitation 
energy. Even the short ranged tensor correlator ($I_{\vartheta}^{(S=1,T=0)}$=0.07 fm$^3$), 
overestimates the experimental values by approx. 8 MeV. By decreasing
the range of the tensor correlator, the quadrupole response
is systematically pushed towards lower energies. In comparison
with the IVGDR, the quadrupole response has a similar dependence on
the range of the tensor correlator. For $^{40}$Ca and the
tensor correlator ranges determined by
$I_{\vartheta}^{(S=1,T=0)}$=0.07, 0.08, and 0.09 fm$^3$, the ISGQR centroid energies read
25.1, 26.2, and 27.1 MeV, respectively. In the cases of heavier nuclei, 
these differences are smaller: for $^{208}$Pb, the centroid
energy lowers by 1.2 MeV when going from the correlator
with $I_{\vartheta}^{(S=1,T=0)}$=0.09 fm$^3$ towards
$I_{\vartheta}^{(S=1,T=0)}$=0.07 fm$^3$.
From a comparative study of low-lying excitations and giant resonances,
we can test the sensitivity of different parts of
the quadrupole transition spectra on the ranges of the tensor
correlator~(Fig.~\ref{figquad2}). Whereas the low-energy $0\hbar\omega$ excitations
depend only weakly on $I_{\vartheta}^{(S=1,T=0)}$, the ISGQR appears to be
more sensitive on variations of the tensor correlator range.

From our results on the ISGMR, IVGDR and ISGQR nuclear response 
we can conclude on the properties of the correlated 
NN interactions, used here as effective interactions in the RPA 
calculations. 
The agreement achieved between the calculated and experimental 
properties of the ISGMR indicates that the correlated NN interaction
corresponds to realistic values of the nuclear matter (NM) incompressibility. 
It has been demonstrated in the past that, within 
relativistic and non-relativistic RPA, 
the energies of the dipole and quadrupole resonances, on one hand, 
and the value of the effective mass corresponding to the effective 
interaction used, on the other, are correlated \cite{Hui.89,Rei.99}.
In particular, the relativistic RPA without density-dependent 
interaction terms, based on the ground state with a small effective
mass and relatively high compression modulus, 
resulted in systematically overestimated energies of 
giant resonances~\cite{Hui.89}.
The discrepancies between UCOM-RPA calculations and experimental
data for multipole giant resonances, as well as the low density
of single-nucleon UCOM-HF states, suggest that the respective
effective mass is too small. These observations are consistent with an exploratory UCOM-HF calculation in NM.

%%%%%%%%%%%%%%%%%%%%%%%%%%%%%%%%%%%%%%%%%%%%%%%%
%  Section VI 

\section{\label{secV}Summary}

In the present study, the correlated interaction $\VO_{\UCOM}$ based on the Argonne V18 potential is employed in fully self-consistent RPA calculations across the nuclear chart. The short-range central and tensor correlations induced by the NN interaction are treated within the Unitary Correlation Operator Method. The precision and self-consistency of the present UCOM-RPA approach are tested in the cases of separation of the spurious center-of-mass motion, and in recovering the sum rules. It is illustrated that this method provides a highly accurate separation of spurious components from the physical spectra, and the sum rules are exhausted up to $\pm3\%$. A consistent implementation of correlated transition multipole operators, even for long-range tensor correlators, results in similar response as the bare operators.

The Hartree-Fock plus UCOM-RPA framework is employed in studies of giant resonances both in light and heavier closed-shell nuclei. The sensitivity of global properties of giant resonances on the range parameters of the tensor correlators are systematically studied. The excitation energies of giant resonances slowly decrease as the range of the tensor correlator is reduced from the standard value $I_{\vartheta}^{(S=1,T=0)}$=0.09 fm$^3$ towards 0.07 fm$^3$.

In comparison to the experimental data, the UCOM-RPA scheme with the standard correlator results in slightly higher resonance energy of the breathing mode ($\approx$ 1-3 MeV), whereas it overestimates the excitation energies of IVGDR and ISGQR by $\approx$ 3-8 MeV. Decreasing of the range of the tensor correlator which improves the description of ISGMR is not sufficient to reproduce experimental data on excitations with higher multipolarities (IVGDR,ISGQR). The increased excitation energies of giant resonances are related to the rather wide HF single particle spectra which are used as a basis for the RPA configuration space, i.e. the small value of the effective mass. Tensor correlations with shorter range increase the single-particle level density, improve the  description of nuclear radii, and result in a systematic shift of the giant resonances towards lower energies.

The correlated realistic interactions are sufficient to generate collective excitation modes, but for an accurate description of experimental data on peak excitation energies and transition strengths, the UCOM-RPA approach should be further extended. In particular this would include the long-range correlations beyond the simple RPA model, i.e., the inclusion of more complex configurations, e.g. the second RPA with coupling of $ph$ with $2p2h$ configurations. The present scheme could also be improved by including the self-consistent coupling to two-particle-hole phonons within the framework of the dressed RPA~\cite{Bar.03}. This approach is advantageous since it accounts for both, the effects of nuclear fragmentation and the RPA correlations due to two-phonon fluctuations in the ground state. 

Apart from long-range correlations, three-nucleon forces might play an
important role for the quantitative description of single-particle
spectra and collective response. Here, we have used correlated
two-nucleon interactions which are able to describe binding energies in
no-core shell-model or many-body perturbation theory without any
supplementary three nucleon force. This is possible due to a
cancellation between the attractive genuine three-nucleon force, which
accompanies the bare Argonne V18 potential, and the repulsive induced
three-body force resulting from the unitary transformation of the
Hamiltonian. This cancellation works surprisingly well for the binding
energies over a large mass range. However, this does not hold for other
observables, e.g. the charge radii. Preliminary results using a
phenomenological three-nucleon force in the Hartree-Fock framework indicate
that radii and single-particle spectra can be improved significantly. As
a consequence an improvement of the calculated excitation energies of
giant resonances with respect to the experimental data can be expected.

The inclusion of the missing correlations and residual three-body interactions in the UCOM-RPA scheme, that should improve the description of low-lying excitations and giant resonances on a quantitative level, will be the subject of our future work. In addition, the UCOM-RPA correlations due to low-lying states and collective excitations, may also provide the long-range correlations which are necessary to improve the HF ground state.

%=========================================================================
\section*{Acknowledgments}

We gratefully acknowledge discussions with Carlo Barbieri and Hans Feldmeier. This work has been supported by
the Deutsche Forschungsgemeinschaft (DFG) under contract SFB 634. We thank the Institute for Nuclear Theory at the University of Washington for its hospitality and the Department of Energy for partial support during the completion of this work.

%==========================================================================

\newpage
\newpage

\end{document}